\documentclass[aps,pra,twocolumn,groupedaddress,showpacs,superscriptaddress,amssymb,amsmath]{revtex4-1}
\usepackage[utf8]{inputenc}
\usepackage{graphicx}
\usepackage{tabularx}
\usepackage{xcolor}
\usepackage{tikz-feynman}
\usepackage{feynmp}
\usepackage{amsmath}

\usepackage{comment}
\usepackage{dcolumn}
\usepackage{hyperref}
\hypersetup{
    colorlinks=true,
    linkcolor=blue,
	citecolor=cyan
}
\usepackage{bm}
\usepackage{epsf}
\usepackage{subfigure}
\usepackage{braket}
\usepackage{tensor}
\usepackage{soul}
\usepackage{float}
\usepackage{mathrsfs}
\usepackage{mathtools}
\usepackage{multirow}
\usepackage{caption}
\usepackage{amsfonts}

\DeclareMathOperator{\trace}{Tr}

\newcommand{\be}{\begin{equation}}
\newcommand{\ee}{\end{equation}}
\newcommand{\bea}{\begin{eqnarray}}
\newcommand{\eea}{\end{eqnarray}}

\newcommand\bka[1]{\textcolor{blue}{#1}}

\makeatletter
\newsavebox{\@brx}
\newcommand{\llangle}[1][]{\savebox{\@brx}{\(\m@th{#1\langle}\)}%
  \mathopen{\copy\@brx\kern-0.5\wd\@brx\usebox{\@brx}}}
\newcommand{\rrangle}[1][]{\savebox{\@brx}{\(\m@th{#1\rangle}\)}%
  \mathclose{\copy\@brx\kern-0.5\wd\@brx\usebox{\@brx}}}
\makeatother

\begin{document}
\title{Study of non-equilibrium Green's functions beyond Born approximation in open quantum systems}
\author{Katha Ganguly}
\email{katha.ganguly@students.iiserpune.ac.in}
\affiliation{Department of Physics,
		Indian Institute of Science Education and Research, Pune 411008, India}
\author{Bijay Kumar Agarwalla}
\email{bijay@iiserpune.ac.in}
\affiliation{Department of Physics,
		Indian Institute of Science Education and Research, Pune 411008, India}
\date{\today}
\begin{abstract}
We provide a systematic approach to compute different kinds of non-equilibrium Green's functions for open quantum systems which are essentially two-point correlation functions in time. We reveal that the definition of Green's functions based on the Born approximation does not provide the correct results in the leading order of the system-bath coupling. We next provide a systematic correction term in Green's functions by going beyond the Born approximation and incorporating a finite correlation between the system and the bath. We primarily focus on two paradigmatic models of open quantum systems, namely, the dissipative Caldeira-Leggett model and the dissipative spin-boson model. We show that the inclusion of such a correction correctly reproduces the Kadanoff-Baym type equation for the so-called lesser or greater components of the Green's functions and provides the correct long-time result up to the first non-zero order of the system-bath coupling, satisfying the detailed balance condition in thermal equilibrium. We further extend our study to a system coupled to multiple reservoirs simultaneously that are maintained at different temperatures and obtain expressions for non-equilibrium steady-state energy current, once again correct up to the first non-zero order of the system-bath coupling.
\end{abstract}
  \maketitle  


\section{Introduction}
Green's functions are the central quantities that are often related to experimentally accessible correlators \cite{Schwinger_Brownian,fluctuation-dissipation,NEGF,NEGF_2,NEGF_3}. For non-equilibrium systems, one can usually define six different types of Green's functions, namely the time-ordered, anti-time-ordered, lesser, greater, retarded, and advanced components \cite{Bijay_NEGF,GF, Rammer}. However all these Green's functions are not independent and in fact under the most general scenario, only three Green's functions can be shown to be independent. In contrast, for systems residing in thermal equilibrium, due to the additional fluctuation-dissipation relation \cite{fluctuation-dissipation,R_Kubo_1966,fluct-dissi_1,fluct-dissi_2,fluct-dissi_3}, only one Green's function remains independent \cite{Bijay_NEGF,Rammer}.

In the context of generic open quantum systems, one can compute Green's functions or the two-point correlation functions in various ways, such as following exact numerical techniques \cite{NEGF_numerical}, the non-equilibrium Green's function approach \cite{Bijay_NEGF,negf_l,haug2008quantum, Rammer,Dhar2006,NEGF_6}, or the quantum master equation (QME) approach \cite{Carmichael, breuer,Marco_schiro,QME_1,QME_2,QME_3,QME_4,QME_5,QME_6,Lyapunov}. In particular, in the QME approach,  typical definitions of Green's functions are based on the Born approximation \cite{breuer,dynamics_oqs,Carmichael,Alicki,davies,Gernot}
which relies on decoupling the total density matrix at any instant of time into a product form for the system and the reservoir. Such a decoupling scheme can possibly miss crucial system-reservoir correlation \cite{correlations,correlation_2,Pernice_2012} and can lead to incorrect predictions for Green's functions both in transient as well as in steady-state. Needless to mention, often in addition to such Born approximation, a Markovian approximation \cite{breuer,Bruer_Review,Born-Markov,Weiss,REDFIELD19651,lidar2020lecture} is incorporated that results in the well-known quantum regression theorem (QRT)  \cite{Carmichael, Lindblad, Alicki, breuer,QRT_2,qrt_3,qrt_4,qrt_5,qrt_6} which often lacks the detailed balance condition in thermal equilibrium \cite{detailed_balance,detailed_balance_2,MQRT}. It is, therefore, an important task to understand the fate of the Green's functions under these approximations, including analyzing the well-known thermodynamically consistent relations, such as the detailed balance condition.  

In this work, we focus on different kinds of Green's functions, in particular the so-called the greater and the retarded components. We first start with the greater component and show that its definition under the Born approximation does not produce correct result in the first non-zero order of the system-bath coupling. We then propose a correction to its definition that captures the finite correlation between the system and reservoir which gets completely ignored in the Born approximation. This correction is crucial to receive the correct steady-state results for Green's function, including respecting the Kubo-Martin-Schwinger (KMS) condition \cite{Kubo,KMS,KMS_2,KMS_3} under thermal equilibrium scenario. We further discuss the fate of retarded components under Born approximation. We illustrate our findings in detail for the Caldeira-Leggett model \cite{CALDEIRA1983587,CALDEIRA_2,CL_2,CL_3,breuer} and the spin-boson model \cite{SB_1,SB_2,SB_3,SB_4,SB_5} and show how the correction term leads to a Kadanoff-Baym type equation \cite{Rammer,KB_1,KB_2} for the greater/lesser component of Green's function and provide a correct prediction of the long-time results up to the first non-zero order of the system-bath coupling. We also extend our analysis for the multiple reservoir scenario that supports finite energy current in a non-equilibrium steady-state. 



We organize the paper as follows: In section~\ref{setup}, we provide the details of the setup, the definitions of the Green's function following the Born approximation, and a diagrammatic representation for a particular Green's function component.  In section~\ref{FD-1}, we shed light on the issue with Born approximation and propose a correction term for the Green's function which can produce correct results to the first non-zero order of the system-bath coupling. 
In section~\ref{models}, we provide two paradigmatic model examples -- the dissipative Caldeira-Leggett model and the dissipative spin-boson model, and illustrate how the Born approximated definition fails to provide a thermodynamically consistent long-time result, whereas the correction term helps to get consistent results. In section~\ref{multibath}, we extend our analysis to a multiple reservoir case and compute non-equilibrium steady state current. Lastly, in section~\ref{summary}, we summarize our results. We delegate certain details to the appendix.

\section{Setup and non-equilibrium Green's functions under Born approximation} \label{setup}
Let us consider a general Hamiltonian describing a quantum system of interest that is interacting with a reservoir. We write the total Hamiltonian as  
\begin{equation}
    H=H_{S}+H_{R}+  \lambda \sum_{i=1}^{d} g_{i} S_{i}\otimes R_{i} \label{general_H}.
\end{equation}
Here, $H_{S}$ describes the system Hamiltonian, $H_{R}$ is the Hamiltonian of the reservoir (bath) which is considered to be a quadratic throughout our analysis. The bath is assumed to be consisting of an infinite collection of non-interacting harmonic oscillators. The last term in Eq.~\eqref{general_H} represents the system-reservoir coupling. Here $S_{i}$ is the system operator which is coupled with the bath operators $R_{i}$ and the corresponding coupling parameter is denoted by $g_{i}$. Both $S_{i}$ and $R_{i}$ are chosen to be Hermitian here. The parameter $\lambda$ keeps track of the order of the coupling strength. The summation index $i$  in Eq.~\eqref{general_H} ranges up to $d$ which indicates the number of system and bath operators that are coupled together. 

In this work, we investigate two-point correlation functions or the non-equilibrium Green's functions based on the definition that originates following the Born approximation. 
We first write down the definitions following the Born approximation, given as (we set $\hbar= 1$ throughout the text), 
\begin{align}
    \langle A(t) B(t') \rangle&=\theta(t-t')\mathrm{Tr}_{S}\big[A\hat{\mathcal{V}}(t-t')\big(B\rho_{S}(t')\big)\big]\nonumber\\&+\theta(t'-t)\mathrm{Tr}_{S}\big[B\hat{\mathcal{V}}(t'-t)\big(\rho_{S}(t)A\big)\big]\label{correlator_define},
\end{align}
where $\theta(t)$ is the Heaviside theta function, $\rho_{S}(t)$ is the reduced density matrix of the system at time $t$ and $\hat{\mathcal{V}}(t)$ is a general propagator which is a superoperator here and is denoted by the hat symbol. $\hat{\mathcal{V}}(t)$ describes the time evolution of the reduced density matrix $\rho_{S}$ by $\hat{\mathcal{V}}(t-t_0)\rho_{S}(t_{0})=\rho_{S}(t)$, where we assume $t_0$ as the starting time for the dynamics. Here the Born approximation implies that the total density matrix at any instant of time can be written into a product form of the system and the bath density matrix i.e., $\rho_{\rm tot}(t')=\rho_{S}(t') \otimes \rho_{R}(t')$. As a result of this approximation, one can perform a partial trace  over the bath and finally, the expression is reduced to Eq.~(\ref{correlator_define}) with a trace remaining only over the system (see Appendix \ref{standard_def_proof} for the details). 

One can always relate the two-point correlators with Green's functions. In particular, to study  quantum dynamics, there are various useful kinds of Green's functions one defines following the Non-equilibrium Green's function (NEGF) approach. For example, using Eq.~\eqref{correlator_define}, i.e., following the Born approximation, one can construct greater, lesser, retarded, advanced, time-ordered, and anti-time-ordered Green's functions. The greater component of the Green's function is defined as 
\begin{align}
    G^{>}_{AB}(t,t')=&-i \, \theta(t-t')\mathrm{Tr}_{S}\Big[A\hat{\mathcal{V}}(t-t')\big(B\rho_{S}(t')\big)\big]\nonumber\\&-i \, \theta(t'-t)\mathrm{Tr}_{S}\big\{B\hat{\mathcal{V}}(t'-t)\big(\rho_{S}(t)A\big)\Big].\label{greater_def}
\end{align} 
Similarly, the lesser Green's function can be defined as 
\begin{align}
    G^{<}_{AB}(t,t')&=-i \, \xi \, \theta(t-t')\mathrm{Tr}_{S}\Big[A\hat{\mathcal{V}}(t-t')\big(\rho_{S}(t')B\big)\Big]\nonumber\\&-i\,\xi \, \theta(t'-t)\mathrm{Tr}_{S}\Big[B\hat{\mathcal{V}}(t'-t)\big(A\rho_{S}(t)\big)\Big].\label{lesser_def}
\end{align} 
where $\xi$ can be either $+1$ or $-1$ depending on our choice.

The retarded and advanced Green's functions can be defined as
\begin{align}
    &G^{r}_{AB}(t,t')=-i\,\theta(t\!-\!t')\mathrm{Tr}_{S}\Big[A\hat{\mathcal{V}}(t\!-\!t')\big[B,\rho_{S}(t')\big]_{\xi}\Big],\label{retarded_def} \\
    &G^{a}_{AB}(t,t')\!=- i\xi\theta(t'-t)\mathrm{Tr}_{S}\Big[B\hat{\mathcal{V}}(t'-t)[A, \rho_{S}(t)]_{\xi}\Big].
\end{align} 
where the subscript $\xi=+1$ ($\xi=-1$) corresponds to a commutator (anti-commutator) for $G^{r}$ and $G^{a}$. The time-ordered Green's function can be defined as
\begin{align}
    G^{t}_{AB}(t,t')=&-i\, \theta(t-t')\,\mathrm{Tr}_{S}\Big[A\hat{\mathcal{V}}(t-t')\big(B\rho_{S}(t')\big)\Big]\nonumber\\&-i\xi \theta(t'-t) \, \mathrm{Tr}_{S}\Big[B\hat{\mathcal{V}}(t'-t)\big(A\rho_{S}(t)\big)\Big].
\end{align}
Finally, the anti-time-ordered Green's function can be defined as
\begin{align}
    G^{\bar{t}}_{AB}(t,t')=&-i\xi\,\theta(t-t')\,\mathrm{Tr}_{S}\Big[A\hat{\mathcal{V}}(t-t')\big(\rho_{S}(t')B\big)\Big]\nonumber\\&-i\,\theta(t'-t)\,\mathrm{Tr}_{S}\Big[B\hat{\mathcal{V}}(t'-t)\big(\rho_{S}(t')A\big)\Big].
    \label{G-tbar}
\end{align}
Note that the above Green's functions are not all independent but obey certain relations among themselves. For example,
\begin{align}
     &G^>(t,t') - G^<(t,t')= G^r(t,t') - G^a(t,t'), \\
     &G^t(t,t') - G^{\bar{t}}(t,t') = G^r(t,t')  + G^a(t,t'), \\
          &G^t(t,t') + G^{\bar{t}}(t,t') = G^>(t,t') + G^<(t,t').
\end{align}
As a consequence of the above relations, three of the Green's functions are independent for a generic non-equilibrium system. 
Note that to check whether the above definitions from Born approximation produce consistent results for the dynamics, one can always compare with the exact Green's functions or with exact numerical simulations, whenever possible.
Now to proceed with the above definition [Eq.~\eqref{greater_def}- Eq.~\eqref{G-tbar}], one needs to know the expression of the propagator (super-operator)
$\hat{\mathcal{V}}$. However, obtaining a general propagator $\hat{\mathcal{V}}$ is a highly nontrivial task. Very recently, in Ref.~\cite{Marco_schiro}, a general expression for  $\hat{\mathcal{V}}$ is obtained in the form of a Dyson equation. The central principle for obtaining $\hat{\mathcal{V}}$, as discussed in Ref.~\cite{Marco_schiro}, relies on treating the system-bath interaction Hamiltonian perturbatively while expanding the corresponding evolution operator order by order in the coupling strength $\lambda$, (recall that the parameter $\lambda$ keeps track of the order of the system-bath coupling strength
and is defined via Eq.~\eqref{general_H}) and performing the trace over the reservoir. 
In this work, we are interested in obtaining Green's functions correctly up to $O(\lambda^{2})$. We, therefore, write down the Dyson equation for $\hat{\mathcal{V}}(t)$ to $O(\lambda^{2})$, given as \cite{Marco_schiro}, 
\begin{align}
    &\hat{\mathcal{V}}(t-t_0)=\hat{\mathcal{V}}_{0}(t-t_0)-i\int_{t_{0}}^{t} dt_{1} \int_{t_{0}}^{t_{1}} dt_{2}\hat{\mathcal{V}}_{0}(t-t_{1})\nonumber\\&\sum_{\gamma_{1},\gamma_{2};i,j}\gamma_{1}\gamma_{2} \, \Sigma^{\gamma_{1}\gamma_{2}}_{ij}(t_{1}-t_{2})\,\hat{S}_{\gamma_{1}}^{i}\,\hat{\mathcal{V}}_{0}(t_{1}-t_{2})\,\hat{S}_{\gamma_{2}}^{j}\,\hat{\mathcal{V}}(t_{2}-t_{0}). \label{dyson_V}
\end{align}
Here, $\hat{\mathcal{V}}_{0}(t) \bullet= e^{-i H_S t} \,\bullet \,e^{i H_S t}$ is the bare propagator of the system. The symbol $\gamma_{i}$ in Eq.~\eqref{dyson_V} can take two possible values $+1$ or $-1$. As a result, the second term of Eq.~\eqref{dyson_V} consists of four possible combinations.
The $(i,j)$ indices in the summation describe the different system and bath operators present in the system-bath interaction term in Eq.~\eqref{general_H}.
Also note that the $\hat{S}^{i}_{\gamma_{i}}$ and $\hat{R}^{i}_{\gamma_{i}}$  denotes super-operator form of $S_{i}$ and $R_{i}$ corresponding to sign of  $\gamma_{i}$, as appeared in Eq.~\eqref{dyson_V}. It's operation on an operator is defined as $\hat{S}^{i}_{+}[\bullet]=S^{i}\bullet$ and $\hat{S}^{i}_{-}[\bullet]=\bullet S^{i}$, which also holds for the bath operators $R_{i}$.
Here  $\Sigma^{\gamma_{1}\gamma_{2}}_{ij}(t_{1}-t_{2})$ is the bath self-energy  with its elements defined as,
\begin{align}
    \Sigma^{\gamma_{1}\gamma_{2}}_{ij}(t_{1}-t_{2})=-i\lambda^{2}\,g_{i} \, g_{j}\, \mathrm{Tr}_{R}\big\{\hat{R}_{I,\gamma_{1}}^{i}(t_{1})\hat{R}_{I,\gamma_{2}}^{j}(t_{2})\bar{\rho}_{R}\big\}.
    \label{bath-corr}
\end{align}
where $\bar{\rho}_{R}$ is the thermal Gibbs state of the bath corresponding to the Hamiltonian $H_R$, and as a result, the bath correlation functions are always time-translationally invariant.  
The different components of the self-energy in Eq.~\eqref{bath-corr} arising for the possible combinations of $\gamma_{1},\gamma_{2}$ indices can be arranged in a matrix form as, 
\begin{align} 
\mathbf{\Sigma}(t_{1}-t_{2})&=\begin{bmatrix}
  \Sigma^{++}(t_{1}-t_{2})\,\, &\,\, \Sigma^{+-}(t_{1}-t_{2})\\ 
  \\
  \Sigma^{-+}(t_{1}-t_{2})\,\, &\, \,\Sigma^{--}(t_{1}-t_{2})
\end{bmatrix}\nonumber\\
&=\begin{bmatrix}
  \Sigma^{t}(t_{1}-t_{2})\,\, &\,\, \Sigma^{<}(t_{1}-t_{2})\\ 
  \\
  \Sigma^{>}(t_{1}-t_{2})\,\, &\, \,\Sigma^{\bar{t}}(t_{1}-t_{2})
\end{bmatrix}, \label{bath_self_energy_matrix}
\end{align}
where each entry of the $2d \times 2d$ $\mathbf{\Sigma}$ matrix in  Eq.~\eqref{bath_self_energy_matrix} is itself a $d\times d$ block matrix consisting of correlations defined between $R_{i}$ and $R_{j}$ operators.

Next, we substitute the Dyson equation for the propagator $\hat{\mathcal{V}}$ in Eq.~\eqref{dyson_V}  into the  definition of the greater Green's function in  Eq.~\eqref{greater_def} and we receive the following expression for the correlator  $t \geq t'$, 
\begin{widetext}
   \begin{align}
    G^{>}_{AB}(t,t')&=-i\,\mathrm{Tr}_{S}\Big[A\hat{\mathcal{V}}(t-t')\big(B\hat{\mathcal{V}}(t'-t_0)\rho_{S}(t_0)\big)\Big]\nonumber\\
    &=-i\,\mathrm{Tr}_{S}\Big[A\hat{\mathcal{V}}_{0}(t-t')\big(B\hat{\mathcal{V}}_{0}(t'-t_0)\rho_{S}(t_0)\big)\Big]\nonumber\\&+(-i)^2\int_{t'}^{t}\!\! dt_{1} \!\!\int_{t'}^{t_{1}}\!\! dt_{2}\!\!\!\sum_{\gamma_{1},\gamma_{2};i,j}\!\!\!\gamma_{1}\gamma_{2}\,\Sigma^{\gamma_{1}\gamma_{2}}_{ij}(t_{1}\!\!-\!\!t_{2})\mathrm{Tr}_{S}\Big[A\hat{\mathcal{V}}_{0}(t\!\!-\!\!t_{1})\hat{S}^{i}_{\gamma_{1}}\!\hat{\mathcal{V}}_{0}(t_{1}\!\!-\!\!t_{2})\hat{S}^{j}_{\gamma_{2}}\!\hat{\mathcal{V}}(t_{2}\!\!-\!\!t')\big(B\hat{\mathcal{V}}(t'\!\!-\!\!t_0)\rho_{S}(t_0)\big)\Big]\nonumber\\&+(-i)^2\int_{t_{0}}^{t'}\!\!\! dt_{1}\!\! \int_{t_{0}}^{t_{1}} \!\!\!dt_{2} \!\!\!\sum_{\gamma_{1},\gamma_{2};i,j}\!\!\!\gamma_{1}\gamma_{2}\,\Sigma^{\gamma_{1}\gamma_{2}}_{ij}(t_{1}\!\!-\!\!t_{2})\mathrm{Tr}_{S}\Big[A\hat{\mathcal{V}}_{0}(t\!\!-\!\!t')\big(B\hat{\mathcal{V}}_{0}(t'\!\!-\!\!t_{1})\hat{S}^{i}_{\gamma_{1}}\!\hat{\mathcal{V}}_{0}(t_{1}\!\!-\!\!t_{2})\hat{S}^{j}_{\gamma_{2}}\!\hat{\mathcal{V}}(t_{2}\!\!-\!\!t_{0})\rho_{S}(t_{0})\big)\Big]. \label{diagram_expansion}
  \end{align}
  Note that a similar expression of $G^{>}_{AB}(t,t')$ can be obtained for $t < t'$. We represent the three terms in Eq.~\eqref{diagram_expansion} by Feynman diagrams in Fig.~(\ref{feynman}). The first diagram represents the bare correlator corresponding to the first term in Eq.~\eqref{diagram_expansion}. In Fig.~(\ref{feynman}) we have considered the time evolution from the left to the right. The thin horizontal lines describe free evolution by $\hat{\mathcal{V}}_{0}(t)=e^{-iH_{0}t}\bullet e^{iH_{0}t}$ and 
  the thick horizontal lines describe the time evolution by the Born propagator $\hat{\mathcal{V}}$ in Eq.~\eqref{dyson_V}.
  The solid black circles on the horizontal line represent the time instants when there is a system operator acting.  The dashed semicircle describes the self-energy $\Sigma_{ij}^{\gamma_1 \gamma_2}(t_1-t_2)$.
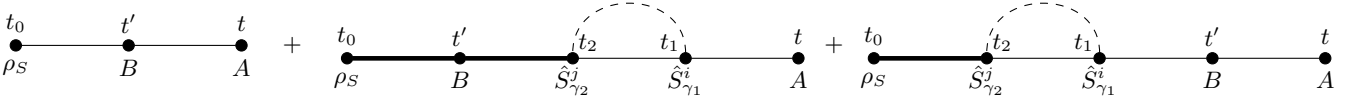
\begin{figure}[h!]
 \[ \vcenter{\hbox{\begin{tikzpicture}
     \begin{feynman}
         \vertex [dot](a){};
         \vertex [right=of a,dot] (b) {};
         \vertex [right=of b,dot] (c) {};
         \diagram{
            (b)--(a);
            (c)--(b);
        };
        \vertex [above=0.9em of a] {$t_{0}$};
        \vertex [below=0.9em of a] {$\rho_{S}$};
        \vertex [above=0.9em of b] {$t'$};
        \vertex [below=0.9em of b] {$B$};
        \vertex [above=0.9em of c] {$t$};
        \vertex [below=0.9em of c] {$A$};
    \end{feynman}
\end{tikzpicture}}} \hspace{0.3cm}+\hspace{0.3cm}
    \vcenter{\hbox{\begin{tikzpicture}
     \begin{feynman}
         \vertex [dot](a){};
         \vertex [right=of a,dot] (b) {};
         \vertex [right=of b,dot] (c) {};
         \vertex [right=of c,dot] (d) {};
         \vertex [right=of d,dot] (e) {};
         \diagram{
            (b)--[ultra thick](a);
            (b)--[ultra thick](a);
            (c)--[ultra thick](b);
            (c)--[ultra thick](b);
            (d)--[scalar,half right](c);
           (d)--(c);
           (e)--(d);
        };
        \vertex [above=0.9em of a] {$t_{0}$};
        \vertex [below=0.9em of a] {$\rho_{S}$};
        \vertex [above=0.9em of b] {$t'$};
        \vertex [below=0.9em of b] {$B$};
        \vertex [above right=0.9em of c] {$t_{2}$};   
        \vertex [below=0.9em of c] {$\hat{S}^{j}_{\gamma_{2}}$};
        \vertex [above left=0.9em of d] {$t_{1}$};    
        \vertex [below=0.9em of d] {$\hat{S}^{i}_{\gamma_{1}}$};
        \vertex [above=0.9em of e] {$t$};
        \vertex [below=0.9em of e] {$A$};
    \end{feynman}
\end{tikzpicture}}} \hspace{0.1cm}+\hspace{0.1cm}
\vcenter{\hbox{\begin{tikzpicture}
    \begin{feynman}
         \vertex [dot](a){};
         \vertex [right=of a,dot] (b) {};
         \vertex [right=of b,dot] (c) {};
         \vertex [right=of c,dot] (d) {};
         \vertex [right=of d,dot] (e) {};
         \diagram{
             (b)--[ultra thick](a);
             (b)--[ultra thick](a);
             (c)--[scalar,half right](b);
             (c)--(b);
             (d)--(c);
             (e)--(d);
         };
        \vertex [above=0.9em of a] {$t_{0}$};
        \vertex [below=0.9em of a] {$\rho_{S}$};
        \vertex [above right=0.9em of b] {$t_{2}$};
        \vertex [below=0.9em of b] {$\hat{S}^{j}_{\gamma_{2}}$};
        \vertex [above left=0.9em of c] {$t_{1}$};
        \vertex [below=0.9em of c]
        {$\hat{S}^{i}_{\gamma_{1}}$};
        \vertex [above=0.9em of d] {$t'$};
        \vertex [below=0.9em of d] {$B$};
        \vertex [above=0.9em of e] {$t$};
        \vertex [below=0.9em of e] {$A$};
     \end{feynman}
\end{tikzpicture}}}
\]
\caption{Diagrammatic representation of the three consecutive terms in Eq.~\eqref{diagram_expansion}. The time evolution is considered from left to right.
The thin horizontal lines describe free evolution by bare propagator $\hat{\mathcal{V}}_{0}$ and the thick horizontal lines describe the time evolution by the Born propagator $\hat{\mathcal{V}}$. The solid black circles on the horizontal line represent the time instants when there is a system operator acting.  Also, the dashed semicircle describes the self-energy due to the bath. The first diagram involves only the bare correlator $\hat{\mathcal{V}}_{0}$. The second and third diagrams are the $O(\lambda^{2})$ contribution of the dissipative bath.} 
\label{feynman}
\end{figure}
\end{widetext}
Following the above description, one can read the diagrams in Fig.~\ref{feynman}, say, for the second diagram,  given an initial state $\rho_{S}(t_{0})$, it evolves by the Born propagator $\hat{\mathcal{V}}$ (thick solid line) till time $t'$ where it becomes $\rho_{S}(t')$ and then the operator $B$ acts (note that we choose here $t \geq t')$.  Following this, the combined operator $(B\rho_{S}(t'))$ evolves by the Born propagator till time $t_{2}$ when the system superoperator $\hat{S}_{\gamma_{2}}^{j}$ operates on this combined operator. After which the combined operator $\big(\hat{S}^{j}_{\gamma_{2}}\!\hat{\mathcal{V}}(t_{2}\!-\!t')\big(B\rho_{S}(t')\big)\big)$ evolves by the free propagator $\hat{\mathcal{V}}_{0}$ till time $t_{1}$ but in presence of system-bath correlation in the form of bath self-energy, as represented by the dashed semi-circle. At time $t_1$ once again the system superoperator $\hat{S}_{\gamma_1}^{i}$ operates on this whole operator $\big(\hat{S}^{j}_{\gamma_{2}}\!\hat{\mathcal{V}}(t_{2}\!-\!t')\big(B\rho_{S}(t')\big)\big)$
Next, the final evolution of the joint operator $\big(\hat{S}^{i}_{\gamma_{1}}\!\hat{\mathcal{V}}_{0}(t_{1}\!-\!t_{2})\hat{S}^{j}_{\gamma_{2}}\!\hat{\mathcal{V}}(t_{2}\!-\!t')\big(B\rho_{S}(t')\big)\big)$ from time $t_1$ to the final time $t$ happens by  the free propagator $\hat{\mathcal{V}}_{0}$ when it finally meets the operator $A$.
This diagram therefore clearly represents the second term in Eq.~\eqref{diagram_expansion}. It is worth noting that the thick line, representing the Born propagator $\hat{\mathcal{V}}$, is important to produce the reducible counterparts of the bath self-energy and appears due to the nature of the Dyson equation, as given in Eq.~\eqref{dyson_V}. The third diagram can be similarly explained and its corresponding expression is represented by the third term in Eq.~\eqref{dyson_V}. A clear difference between the second and the third diagram in Fig.~\eqref{feynman} is that the irreducible part of these diagrams (i.e., replacing $\hat{\mathcal{V}}$ by $\hat{\mathcal{V}}_{0}$), contain system-bath correlation (self-energy) between time interval $[t', t]$ and $[t_0, t]$, respectively. A crucial observation from the irreducible part of the diagrams that follow from this analysis is that the system-bath correlation always appears after (second diagram) and before (third diagram) the time $t'$ and no system-bath correlation is taken into consideration at the time $t'$ and this is justified by the Born approximation. In what follows, we will first put forward the correction diagram and later exemplify via model examples the importance of this diagram in providing a consistent physical picture. 
\vspace{0.2cm}
\section{Correction to two-point correlators beyond Born approximation}
\label{FD-1}
As discussed in the previous section, the crucial Born or the decoupling approximation for the density matrix i.e., at $t'$, $\rho(t')=\rho_{S}(t')\otimes \bar{\rho}_{R}$, leads to an important missing Feynman diagram. This missing diagram in fact captures the finite correlation between the system and the bath at time $t'$ in $O(\lambda^{2})$. We present this new Feynman diagram in Fig.~(\ref{Correction}) where importantly the bath correlation function or the self-energy remains active within the time window $t_1$ and $t_2$, encircling the time $t'$, thereby crucially reflecting the system-bath correlation at time $t'$. In what follows, we will show that capturing this new diagram provides exact answers to the Green's functions up to $O(\lambda^{2})$. 
We, therefore, write the modified definition of the greater Green's function (denoted by $\overline{G}^{>}_{AB}(t,t')$) for $t\ge t'$, obtained by following the Born approximated standard definition in Eq.~\eqref{correlator_define} and the additional correction in Eq.~\eqref{general_cor} as the following, 
\begin{equation}
\overline{G}^{>}_{AB}(t,t') \!=\!G^{>}_{AB}(t,t')+G_{c,AB}^{>}(t,t'),\label{modified_correlator}
\end{equation}
where the last term represents the correction term $[G_{c,AB}^{>}(t,t')]$ due to the new Feynman diagram in Fig.~(\ref{Correction}). Using the diagrammatic rules mentioned earlier in Sec.~\ref{setup}, the mathematical expression of this new diagram  can be written down in the form,
%
%

\begin{widetext}
 \begin{center}
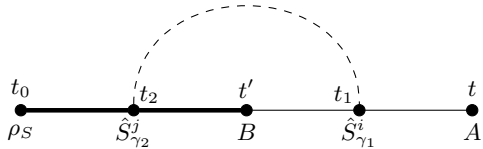
\begin{figure}[h!]
 
\[
\vcenter{\hbox{\begin{tikzpicture}
    \begin{feynman}
         \vertex [dot](a){};
         \vertex [right=of a,dot] (b) {};
         \vertex [right=of b,dot] (c) {};
         \vertex [right=of c,dot] (d) {};
         \vertex [right=of d,dot] (e) {};
         \diagram{
             (b)--[ultra thick](a);
             (b)--[ultra thick](a);
             (b)--[scalar,half left](d);
             (c)--[ultra thick](b);
             (c)--[ultra thick](b);
             (d)--(c);
             (e)--(d);
         };
        \vertex [above=0.9em of a] {$t_{0}$};
        \vertex [below=0.9em of a] {$\rho_{S}$};
        \vertex [above right=0.9em of b] {$t_{2}$};
        \vertex [below=0.9em of b] {$\hat{S}^{j}_{\gamma_{2}}$};
        \vertex [above=0.9em of c] {$t'$};
        \vertex [below=0.9em of c] {$B$};
        \vertex [above left=0.9em of d] {$t_{1}$};
        \vertex [below=0.9em of d] {$\hat{S}^{i}_{\gamma_{1}}$};
        \vertex [above=0.9em of e] {$t$};
        \vertex [below=0.9em of e] {$A$};
     \end{feynman}
\end{tikzpicture}}}
\]
\caption{Feynman diagram that captures the system-reservoir correlation of $O(\lambda^{2})$ at time $t'$ via exchange of energy quanta at time $t_1$ and $t_2$  encircling the time $t'$.  Such a diagram is absent in the standard definition of Green's functions that follows from the Born approximation. Here the dashed semi-circle corresponds to the self-energy due to the bath which encircles the time $t'$ which is important to get the correct Green's function at $O(\lambda^{2})$. }
\label{Correction}
\end{figure}
 \end{center}
    \begin{align}
        G^{>}_{c,AB}&= (-i)^{2}\int_{t'}^{t}\!\!\!dt_{1}\!\!\int_{t_{0}}^{t'}\!\!\!dt_{2}\!\!\sum_{\gamma_{1},\gamma_{2};i,j}\!\!\!\gamma_{1}\gamma_{2}\,\Sigma^{\gamma_{1}\gamma_{2}}_{ij}(t_{1}\!\!-\!\!t_{2})\mathrm{Tr}_{S}\Big[A\hat{\mathcal{V}}_{0}(t\!\!-\!\!t_{1})\hat{S}^{i}_{\gamma_{1}}\!\hat{\mathcal{V}}_{0}(t_{1}\!\!-\!\!t')B\hat{\mathcal{V}}(t'\!\!-\!\!t_{2})\hat{S}^{j}_{\gamma_{2}}\!\hat{\mathcal{V}}(t_{2}\!\!-\!\!t_{0})\rho_{S}(t_{0})\Big],
        \label{general_cor}
    \end{align}
\end{widetext}
where note that the subscript $c$ refers to the correction term. In Appendix \ref{Dyson_eqn_G_prrof_NEGF}, we illustrate that adding this new Feynman diagram in Fig.~\ref{Correction} to the Born approximated standard definition of greater Green's function in Eq.~\eqref{greater_def} results in the modified greater component $\overline{G}^{>}_{AB}(t,t')$ correct up to order $O(\lambda^2)$. 

It is important to note that the diagram in Fig.~\ref{Correction} is very different in nature compared to the Feynman diagrams in Fig.~\ref{feynman} that follow from the Born approximated definition. In particular, the self-energy or the bath-bath correlation function is often related to the exchange of energy quanta between the system and the bath. Such self-energy in Fig.~\ref{feynman} appears either before or after the observation time $t'$, whereas, in Fig.~\ref{Correction} the self-energy encircles the observation time $t'$ and therefore a finite correlation builds up between the system and the bath via the exchange of energy quanta at time $t'$ which is completely absent in Fig.~\ref{feynman}.

In the later sections, we will focus on two paradigmatic open quantum system models, namely the dissipative harmonic oscillator setup and a dissipative spin-boson setup to show the impact of this correction term and how it helps to obtain results correct up to $O(\lambda^2)$. Needless to mention, our approach can be systematically generalized for extended systems. Note that similar corrections like the above can be obtained for other kinds of Green's functions such as lesser, retarded, or advanced components. In what follows, we will now consider model examples and illustrate in detail the consequences of Born approximation and the importance of the new Feynman diagram in the context of non-equilibrium Green's functions.

\section{Model examples} \label{models}
\subsection{Dissipative harmonic oscillator -- Caldeira-Leggett model}
As a first example, we consider the Caldeira-Leggett (CL) model which can be exactly solved following either the quantum Langevin equation approach or the NEGF technique. We first write down the total Hamiltonian for the CL model as, 
\begin{align}
    H=\frac{p^{2}}{2M}\!+\!\frac{1}{2}k_{0}x^{2}\!+\!k'x x_{1}\!+\!\sum_{n}\Big(\frac{p^{2}_{n}}{2m}\!+\!\frac{1}{2}k'_{n}(x_{n}-x_{n+1})^{2}\!\Big),\label{HO_Ham}
\end{align}
where the first two terms in the above Hamiltonian correspond to the system $H_{S}$ for a single harmonic oscillator of mass $M$ and force constant $k_0$. The third term represents the coupling between the system and the bath coordinates with coupling strength $k'$.  The last term represents the Hamiltonian for the reservoir $H_R$, consisting of an infinite collection of coupled harmonic oscillators with force constant $k_n'$. Needless to mention, although we are focusing here on a single oscillator problem, our analysis presented below can be systematically generalized to an extended system such as a coupled harmonic lattice chain \cite{Goold2019,dvira}.  

 Let us first define the greater component of Green's functions for this model using the definition that follows from the Born approximation i.e., Eq.~\eqref{greater_def}. 
For $t\ge t'$, the greater component for the CL model is defined as 
\begin{align}
   &G^{>}(t,t')=-i\,\mathrm{Tr}_{S}\Big[x \, \hat{\mathcal{V}}(t-t')\,(x\rho_{S}(t'))\Big].
   \label{greater_old_def}
\end{align}
Our task now is to obtain the equation of motion (EOM) for this Green's function. For this purpose, we rely on the EOM of $\hat{\mathcal{V}}(t-t')$ which for this model can be expressed as,
\begin{align}
    \frac{\partial}{\partial t}& \hat{\mathcal{V}}(t-t')=\hat{\mathcal{H}}_{S}\hat{\mathcal{V}}(t-t')\!-i\!\!\int_{t'}^{t} dt_1 \sum_{\gamma_{1},\gamma_{2}}\!\!\gamma_1\gamma_2 \nonumber\\&\Sigma^{\gamma_{1}\gamma_{2}}(t-t_1) \hat{x}_{\gamma_{1}} \hat{\mathcal{V}}_0(t-t_1)\hat{x}_{\gamma_{2}}\hat{\mathcal{V}}(t_1-t_0).
    \label{nu-EOM-CL}
\end{align}
Here $\hat{\mathcal{H}}_{S}$ is the system Hamiltonian superoperator and $\hat{x}_{\gamma_1}\, ,\, \hat{x}_{\gamma_2}$ are the superoperator form of the system operator $x$ that is coupled with the reservoir operator $x_{1}$, as follows from Eq.~\eqref{HO_Ham}. Also $\Sigma^{\gamma_1\gamma_2}(t-t_1)$ is the self-energy due to the reservoir which is given by,
\begin{align}
    \Sigma^{\gamma_1\gamma_2}(t-t_1)=-i\,k'^{2}\,\mathrm{Tr}_{R}\Big[\hat{x}^{I}_{1\,\gamma_1}(t)\hat{x}^{I}_{1\,\gamma_2}(t_1)\bar{\rho}_{R}\Big], \label{bath_self_energy}
\end{align}
where the superscript $I$ in Eq.~\eqref{bath_self_energy} corresponds to the interaction picture representation of the bath operators $x_1$.
Using Eq.~\eqref{nu-EOM-CL}, we receive a closed second-order EOM for the $G^>(t,t')$. This can be seen as follows:
\begin{align}
    \frac{\partial}{\partial t} G^{>}(t,t')&=-i\,\mathrm{Tr}_{S}\big[x \, \partial_t \hat{\mathcal{V}}(t-t')\,(x\rho_{S}(t'))\big]\\
    &=(-i)^2\,\mathrm{Tr}_{S}\big[[x,H_S]\, \hat{\mathcal{V}}(t-t')\,(x\rho_{S}(t'))\big]\\
    &=-\,\frac{i}{M}\mathrm{Tr}_{S}\big[p \, \hat{\mathcal{V}}(t-t')\,(x\rho_{S}(t'))\big].
\end{align}
For simplicity, we choose $M=1$. Note that, the second term in Eq.~\eqref{nu-EOM-CL} does not contribute to the first derivative of $G^{>}(t,t')$. Now taking another derivative with respect to the time $t$, we receive,
\begin{align}
    \frac{\partial^2}{\partial t^2} G^{>}(t,t')&=-i\,\mathrm{Tr}_{S}[p \, \partial_t \hat{\mathcal{V}}(t-t')\,(x\rho_{S}(t'))]\\
    &=i\,k_0\mathrm{Tr}_{S}[x \, \hat{\mathcal{V}}(t-t')\,(x\rho_{S}(t'))]\nonumber\\&+(-i)^{2}\!\sum_{\gamma_{1},\gamma_{2}}\!\!\gamma_1\gamma_2 \int_{t'}^{t}\!\! dt_1 \Sigma^{\gamma_{1}\gamma_{2}}(t\!\!-\!\!t_1) \nonumber\\&\mathrm{Tr}_{S}\Big[p\,\hat{x}_{\gamma_{1}}\! \hat{\mathcal{V}}_0(t\!\!-\!\!t_1)\hat{x}_{\gamma_{2}}\!\hat{\mathcal{V}}(t_1\!\!-\!\!t')(x\rho_S(t'))\Big].
\end{align}
Simplifying the above expression, we finally receive 
\begin{equation}
    \frac{\partial^{2}G^{>}(t,t')}{\partial t^{2}}\!=\!-k_{0}G^{>}(t,t')\!-\!\int_{t'} ^{t}\!\!\! dt_{1} \Sigma^{r}(t\!-\!t_{1}) G^{>}(t_{1},t'),\label{greater_eom}
\end{equation}
where recall that $\Sigma^r$ is the retarded self-energy due to the bath and following Eq.~\eqref{bath_self_energy} is given by  
\begin{equation}
\Sigma^r(t-t') = -i \theta(t-t') k'^2 \big\langle \big[x^{I}_{1}(t), x^{I}_{1}(t')\big]\big\rangle.
\end{equation}
On the other hand, the exact EOM for the greater component [$G_{\rm ex}^{>}(t,t')$] can be obtained for this model following the quantum Langevin equation or the exact NEGF approach and is given by the so-called Kadanoff-Baym equation \cite{Rammer,Bijay_NEGF} 
\begin{align}
    \frac{\partial^{2}G_{\rm ex}^{>}(t,t')}{\partial t^{2}}&=-k_{0}G_{\rm ex}^{>}(t,t')-\int_{t_0}^{t} dt_{1} \Big[\Sigma^{r}(t-t_{1}) G_{\rm ex}^{>}(t_{1},t')\nonumber\\&\quad \quad +\Sigma^{>}(t-t_{1}) G_{\rm ex}^{a}(t_{1},t')\Big]. \label{greater_eom_standard}
\end{align}
Comparing Eq.~\eqref{greater_eom} and the exact EOM in Eq.~\eqref{greater_eom_standard}, we observe that a
crucial inhomogenous term  $\Sigma^> G^{a}$ is not captured in Eq.~\eqref{greater_eom} i.e., at the Born approximation level. Also note that the limit of integration in Eq.~\eqref{greater_eom_standard} is from the starting time of the dynamics $t_0$ to the final time $t$, instead of $t'$ to $t$, as predicted by the Born correlator in Eq.~\eqref{greater_eom}.
Interestingly, both the issues in Eq.~\eqref{greater_eom_standard} get fixed by the new Feynman diagram, as presented in Sec.~\ref{FD-1} which takes into account the finite correlation between system and bath at time $t'$. Needless to mention, the appearance of the inhomogeneous term  $\Sigma^> G^{a}$ via this new diagram plays a pivotal role in getting a physically consistent long-time equilibrium result for the model. In other words, the time dynamics that follows from Eq.~\eqref{greater_eom} do not capture the correct physics for this model. In particular, Eq.~\eqref{greater_eom} contains only the retarded component of the self-energy $\Sigma^r$ which is temperature independent for the bosonic bath, and hence detailed balance condition is never satisfied. 


For this model, following Eq.~\eqref{general_cor}, the correction term $[G_{c}^{>}(t,t')]$ is given by (for $t \geq t'$), 
\begin{widetext}
\begin{align}\label{G>_CL}
    G_{c}^{>}(t,t')&=(-i)^{2}\int_{t'}^{t}dt_{1}\int_{t_{0}}^{t'} dt_{2}\sum_{\gamma_{1}\gamma_{2}}\gamma_{1}\gamma_{2}\,\Sigma^{\gamma_{1}\gamma_{2}}(t_{1}-t_{2})\mathrm{Tr}_{S}\Big[x\hat{\mathcal{V}}_0(t-t_{1})\hat{x}_{\gamma_{1}}\hat{\mathcal{V}}_0(t_{1}-t')x\hat{\mathcal{V}}(t'-t_{2})\hat{x}_{\gamma_{2}}\rho_{S}(t_{2})\Big].
\end{align}  
As a result, finally, the modified greater Green's function, correct up to order $O(\lambda^{2})$ becomes,
\begin{align}
    \overline{G}^{>}(t,t')=G^{>}(t,t')+G_{c}^{>}(t,t'), \label{greater_def_corrected}
\end{align}
where recall that the first term follows from the Born approximation and the second term is the correction from the new Feynman diagram. To obtain the EOM of the modified greater Green's function, we first obtain the EOM of $G^{>}_{c}(t,t')$ in Eq.~\eqref{G>_CL} as the following,
\begin{align}
    \frac{\partial}{\partial t} G^{>}_{c}(t,t')&=(-i)^2 \int_{t_0}^{t'} dt_2 \sum_{\gamma_1\gamma_2} \gamma_1\gamma_2 \Sigma^{\gamma_1\gamma_2}(t-t_2) \mathrm{Tr}_{S}\Big[x\,x_{\gamma_1}\hat{\mathcal{V}}_0(t_{1}-t')x\hat{\mathcal{V}}(t'-t_{2})\hat{x}_{\gamma_{2}}\rho_{S}(t_{2})\Big]\nonumber\\& + (-i)^2 \int_{t'}^{t}dt_{1}\int_{t_{0}}^{t'} dt_{2}\sum_{\gamma_{1}\gamma_{2}}\gamma_{1}\gamma_{2}\Sigma^{\gamma_{1}\gamma_{2}}(t_{1}-t_{2})\mathrm{Tr}_{S}\Big[x\partial_t\hat{\mathcal{V}}_0(t-t_{1})\hat{x}_{\gamma_{1}}\hat{\mathcal{V}}_0(t_{1}-t')x\hat{\mathcal{V}}(t'-t_{2})\hat{x}_{\gamma_{2}}\rho_{S}(t_{2})\Big]\\
    &=(-i)^{2}\int_{t'}^{t}dt_{1}\int_{t_{0}}^{t'} dt_{2}\sum_{\gamma_{1}\gamma_{2}}\gamma_{1}\gamma_{2}\Sigma^{\gamma_{1}\gamma_{2}}(t_{1}-t_{2})\mathrm{Tr}_{S}\Big[p\hat{\mathcal{V}}_0(t-t_{1})\hat{x}_{\gamma_{1}}\hat{\mathcal{V}}_0(t_{1}-t')x\hat{\mathcal{V}}(t'-t_{2})\hat{x}_{\gamma_{2}}\rho_{S}(t_{2})\Big].
\end{align}
Now, taking another derivative with respect to time $t$, we obtain,
\begin{align}
    \frac{\partial^{2}}{\partial t^{2}} G^{>}_{c}(t,t')&=(-i)^2 \int_{t_0}^{t'} dt_2 \sum_{\gamma_1\gamma_2} \gamma_1\gamma_2 \Sigma^{\gamma_1\gamma_2}(t-t_2) \mathrm{Tr}_{S}\Big[p\,x_{\gamma_1}\hat{\mathcal{V}}_0(t_{1}-t')x\hat{\mathcal{V}}(t'-t_{2})\hat{x}_{\gamma_{2}}\rho_{S}(t_{2})\Big]\nonumber\\& + (-i)^2 \int_{t'}^{t}dt_{1}\int_{t_{0}}^{t'} dt_{2}\sum_{\gamma_{1}\gamma_{2}}\gamma_{1}\gamma_{2}\Sigma^{\gamma_{1}\gamma_{2}}(t_{1}-t_{2})\mathrm{Tr}_{S}\Big[p\,\partial_t\hat{\mathcal{V}}_0(t-t_{1})\hat{x}_{\gamma_{1}}\hat{\mathcal{V}}_0(t_{1}-t')x\hat{\mathcal{V}}(t'-t_{2})\hat{x}_{\gamma_{2}}\rho_{S}(t_{2})\Big]\\
    &=i \int_{t_0}^{t'}\!\!\! dt_1 \Big\{\Sigma^{t}(t-t_1)\trace_{S}\Big[x\hat{\mathcal{V}}(t'-t_1)x\rho_S(t_1)\Big]\!\!-\!\!\Sigma^{<}(t-t_1)\trace_{S}\Big[x\hat{\mathcal{V}}(t'-t_1)\rho_S(t_1)x\Big]\Big\}-k_0 G^{>}_{c}(t,t')\\
    &=-k_0 G^{>}_{c}(t,t')-\int_{t_0}^{t'} dt_1 \big[\Sigma^{>}(t-t_1)G^{<}(t_1,t')-\Sigma^{<}(t-t_1)G^{>}(t_1,t')\big].\label{final_G_c_greater_eom}
\end{align}
Here we have used the relation $\Sigma^{t}(t-t_1)=\Sigma^{>}(t-t_1)$ for $t\ge t_1$. Now simplifying the RHS of Eq.~\eqref{final_G_c_greater_eom}, we obtain the EOM of $G^{>}_{c}(t,t')$ as the following,
\begin{align}
    \frac{\partial^2}{\partial t^2}G^{>}_{c}(t,t')=-k_0 G^{>}_{c}(t,t') - \int_{t_0}^{t'} dt_1& \Big[ \Sigma^{>}(t-t_1)G^{a}(t_1,t')+\Sigma^{r}(t-t_1)G^{>}(t_1,t')\Big]. \label{Gc_eom}
\end{align}
\end{widetext}
Interestingly, the EOM of the correction term $G^{>}_{c}(t,t')$ captures the important inhomogeneous term $\Sigma^{>}G^{a}$. The appearance of $\Sigma^{>}$ component of the self-energy reveals the exchange of energy quanta between the system and the bath, and thereby capturing a finite correlation as discussed in Sec.~\ref{FD-1}. Now, adding the EOM in Eq.~\eqref{greater_eom} obtained from Born approximated definition and the EOM of the correction term $G^{>}_{c}(t,t')$ in Eq.~\eqref{Gc_eom}, we receive the EOM of the modified greater Green's function as the following,
\begin{align}
    \frac{\partial^{2}}{\partial t^{2}}\overline{G}^{>}(t,t')\!=\!&-k_{0} \overline{G}^{>}(t,t')\!-\!\!\int_{t_{0}}^{t}\!\! dt_{1} \Sigma^{r}(t,t_{1})\overline{G}^{>}(t_{1},t')\nonumber\\&-\int_{t_{0}}^{t}\!\! dt_{1}\Sigma^{>}(t,t_{1}) G^{a}(t_{1},t')+O(\lambda^4).\label{greater_eom_corrected}
\end{align}
Note that, the EOM of the modified greater Green's function in Eq.~\eqref{greater_eom_corrected} is similar to the Kadanoff-Baym equation in Eq.~\eqref{greater_eom_standard} up to order $O(\lambda^{2})$. Our next aim is to obtain the steady state solution of the greater component of Green's function for this model correct up to $O(\lambda^{2})$. In Appendix \ref{proof_correct_long_time_GF}, we provide a rigorous proof that given a $O(\lambda^{2})$ correct equation of motion for the Green's function, always results in $O(\lambda^{2})$ correct long-time solution. We therefore use the EOM in Eq.~\eqref{greater_eom_corrected} and obtain the steady state solution. 


Note that, in the steady state, all the Green's functions in Eq.~\eqref{greater_eom_corrected} have time-translational invariance. Therefore, considering the long time limit (steady state limit) in Eq.~\eqref{greater_eom_corrected} i.e., $t \to \infty, t' \to \infty$ and keeping $t-t'$ finite and finally performing Fourier transformation with respect to $t-t'$, we obtain
\begin{align}
    \Big[-\omega^{2}+k_{0}+&\Sigma^{r}(\omega)\Big]
    \overline{G}^{>}(\omega)=-\Sigma^{>}(\omega)G^{a}(\omega),\nonumber\\
    &\overline{G}^{>}(\omega)=G^{r}(\omega)\Sigma^{>}(\omega)G^{a}(\omega). \label{long_time_greater_CL}
\end{align}
where $G^r(\omega) = \Big[\omega^{2}-k_{0}-\Sigma^{r}(\omega)\Big]^{-1}$ and $G^a(\omega)=\big[G^r(\omega)\big]^{*}$. Here we use the Fourier transform convention, 
\begin{equation}
f(\omega) = \int_{-\infty}^{\infty}\!\! dt\,\, e^{i \omega t} f(t).
\end{equation}
Remarkably, we receive the exact long-time result for $G^{>}(\omega)$ for this model in Eq.~\eqref{long_time_greater_CL} which is correct in all orders of system-bath coupling. This is due to the fact that, as the $\Sigma^{>}(t,t_{1})G^{a}(t_{1},t')$ in Eq.~\eqref{greater_eom_corrected} 
does not get corrected even in higher orders of system-bath coupling, the long-time solution turns out to be exact. However, note that this happens also due to an important reason that both $G^r$ and $G^a$ Green's functions turn out to be exact for this model, already at the Born correlator level, as we will show later. Note that, exactly similar steps can be followed to obtain the long-time solution for $G^{<}(\omega)$, and is given as 
\begin{equation}
\overline{G}^{<}(\omega)=G^{r}(\omega)\Sigma^{<}(\omega)G^{a}(\omega).
\label{G-lesser}
\end{equation}
The obtained solutions for $\overline{G}^{<}(\omega)$ and $\overline{G}^{>}(\omega)$ satisfy the detailed balance condition and can be seen as follows.  First, the form of the greater and lesser component of bath 
self-energy for the bosonic bath is given by
\begin{align}
&\Sigma^{>}(\omega)=-i\,J(\omega)\big(1+n_{B}(\omega)\big),\\
&\Sigma^{<}(\omega)=-i\,J(\omega)n_{B}(\omega).
\end{align}
Here, $n_{B}(\omega)=1/(e^{\beta \hbar \omega}-1)$ is the Bose-Einstein distribution function and $J(\omega)$ is the bath spectral density which is of the form, 
\begin{equation} 
J(\omega)=2\pi\sum_k |h_k|^2 \delta(\omega-\omega_k),
\end{equation} 
where $\omega_k$ is the eigen-frequency for the $k$-th eigen-mode corresponding to the bath Hamiltonian, given in Eq.~\eqref{HO_Ham} and  $h_k$ is the coupling strength between the k-th eigen-mode and the system operator $x$. It is easy to see that the greater and the lesser components of the bath self-energy satisfy the detailed balance condition $\Sigma^{>}(\omega)=e^{\beta \omega} \Sigma^{<}(\omega)$. Therefore following Eq.~\eqref{long_time_greater_CL} and Eq.~\eqref{G-lesser}, it is obvious that $\overline{G}^{>}(\omega)$ and $\overline{G}^{<}(\omega)$ will also satisfy the same detailed balance or the condition, i.e.,  $\overline{G}^{>}(\omega)=e^{\beta \omega}\overline{G}^{<}(\omega)$. 
In summary, we see that the Born approximated Green's function does not provide the correct long-time solution including violating the detailed balance, whereas, our proposed correction rectifies the problem.

Next, similar to the greater component, we analyze the retarded and advanced Green's functions for the CL model. Following the Born approximation the retarded component is defined as (we consider $\xi = +1$ in Eq.~\eqref{retarded_def}),
\begin{equation}
G^r(t,t')=-i\,\theta(t-t')\mathrm{Tr}_{S}\Big[x\hat{\mathcal{V}}(t-t')\big[x,\rho_{S}(t')\big]\Big], \label{retarded_CL_def}
\end{equation}
and the advanced component is given as,
\begin{equation}
G^a(t,t')=-i\,\theta(t'-t)\mathrm{Tr}_{S}\Big[x\hat{\mathcal{V}}(t'-t)\big[x,\rho_{S}(t)\big]\Big].
\end{equation}
We primarily focus on the retarded component and the advanced component can be similarly obtained.  As before, we can obtain the EOM of the retarded Green's function in Eq.~\eqref{retarded_CL_def} which is given by,
\begin{align}
    \frac{\partial^{2}G^{r}(t,t')}{\partial t^{2}}&=-\delta(t-t')-k_{0}G^{r}(t,t')\nonumber\\&-\int_{t_{0}}^{t} dt_{1}\Sigma^{r}(t-t_{1}) G^{r}(t_{1},t'). \label{retarded_eom}
\end{align}
Interestingly, we observe that for the CL model, the Born approximated retarded correlator produces the exact EOM for $G^{r}(t,t')$ (see Appendix-\ref{exact_dyson_Gr} for details), and hence for this model it does not require any correction. This is due to the quadratic or non-interacting nature of the setup. In Appendix-\ref{Gr_corr_zero}, we provide a general condition when the Born approximated retarded correlator for a generic setup does not need any correction in $O(\lambda^2)$.

We then immediately receive the correct long-time solution in the frequency domain and is given as 
\begin{align}
    G^{r}(\omega)&=\big[\omega^{2}-k_{0}-\Sigma^{r}(\omega)\big]^{-1}.
\end{align}
Here $\Sigma^{r}(\omega)$, is the retarded component of the self-energy due to the bath. In a similar way, one can obtain for the advanced Green's function the correct long-time solution, given as, 
\begin{align}
     G^{a}(\omega)&=\big[\omega^{2}-k_{0}-\Sigma^{a}(\omega)\big]^{-1} =[G^{r}(\omega)]^{*} ,
\end{align}
where $\Sigma^{a}(\omega)=\big(\Sigma^r(\omega)\big)^{*}$, is the advanced component of the self-energy due to the bath.

In summary, we find that the standard definition following Born approximation works perfectly for the CL model for retarded and advanced Green's function whereas fails drastically for the greater or lesser components and to receive the correct long-time solution, a correction is needed.  Having discussed the CL model in detail, we now shift our focus to another paradigmatic open quantum system model, namely the dissipative spin-boson model. 
%
%
%
%
\subsection{Dissipative spin-boson model}
The Hamiltonian for the dissipative spin-boson (SB) model is given as, 
\begin{equation}
    H=\frac{\omega_{0}}{2}\sigma_{z}+\sum_{k} \omega_{k}b_{k}^{\dagger}b_{k}+\sigma_{x}\sum_{k}g_{k}(b_{k}+b_{k}^{\dagger}),
\end{equation}
where the first term represents a single spin/qubit with the energy difference between its excited and ground state as $\omega_0$ and the spin is in contact with a dissipative bosonic bath. The second term represents the usual bath Hamiltonian and the last term represents the coupling between the spin and the bath. 

We once again start with the standard definition of the greater Green's function based on the Born approximation. For this setup, it is possible to construct nine different combinations of greater Green's function in various combinations of $\sigma_x$, $\sigma_y$ and $\sigma_z$ operators. In matrix form, we write, 
\begin{align}
    G^{>}(t,t')=\begin{pmatrix}
        G^{>}_{xx}(t,t') & \, G^{>}_{xy}(t,t') & \, G^{>}_{xz}(t,t') \\
        \\
        G^{>}_{yx}(t,t') &\, G^{>}_{yy}(t,t') & \,G^{>}_{yz}(t,t') \\
        \\
        G^{>}_{zx}(t,t') &\, G^{>}_{zy}(t,t') \,& G^{>}_{zz}(t,t') \\
    \end{pmatrix}.
\end{align}
Let us first start with the $G^{>}_{xx}$ and $G^{>}_{yx}$ components, as we will see that their EOM gets closed for $t \geq t'$. As before, the standard definition of $G^{>}_{xx}(t,t')$ and $G^{>}_{yx}(t,t')$ for $t\geq t'$ is
\begin{align}
    G^{>}_{xx}(t,t')&=-i\,\mathrm{Tr}_S \Big[\sigma_{x}\hat{\mathcal{V}}(t-t')(\sigma_{x}\rho_{S}(t'))\Big], \label{greater_xx_sb_old}\\
    G^{>}_{yx}(t,t')&=-i\,\mathrm{Tr}_S \Big[\sigma_{y}\hat{\mathcal{V}}(t-t')(\sigma_{x}\rho_S(t'))\Big].\label{greater_yx_sb_old}
\end{align}
The corresponding EOMs for $G^{>}_{xx}(t,t')$ and $G^{>}_{yx}(t,t')$ can be obtained, similarly like the CL model, by writing down the EOM for 
 $\hat{\mathcal{V}}(t-t')$. We obtain the EOMs as, 
 \begin{align}
    \label{Gxx-SB}
    &\frac{\partial}{\partial t}G^{>}_{xx}(t,t')\!=\!-\omega_{0}\,G^{>}_{yx}(t,t'),\\
    &\frac{\partial}{\partial t}G^{>}_{yx}(t,t')=\omega_{0}\,G^{>}_{xx}(t,t')\!-\!2\,i \!\int_{t'}^{t}\!\!\! dt_{1} \Sigma^{K}(t,t_{1})G^{>}_{yx}(t_{1},t'),
    \label{Gyx-SB}
\end{align}
where we introduce $\Sigma^{K}(t,t')=\Sigma^{>}(t,t')+\Sigma^{<}(t,t')$ as the Keldysh component of the bath self-energy. Here $\Sigma^{>}(t,t')$ and $\Sigma^{<}(t,t')$ are defined as,
\begin{align}
    \Sigma^{>}(t,t')=-i\sum_{k} |g_k|^2  \trace_{R}\Big[R^{I}_{k}(t)R^{I}_{k}(t')\bar{\rho}_{R}\Big],\\
    \Sigma^{<}(t,t')=-i\sum_{k} |g_k|^2 \, \trace_{R}\Big[R^{I}_{k}(t')R^{I}_{k}(t)\bar{\rho}_{R}\Big],
\end{align}
where $R_{k}$ is defined as $R_{k}=(b_{k}+b^{\dagger}_{k})$ and the superscript $I$ 
here corresponds to the interaction picture representations of the operators $R_k$.
Once again, we observe that the above coupled differential equations in Eqs.~\eqref{Gxx-SB} and \eqref{Gyx-SB} do not capture any inhomogeneous term as typically expected to obtain a correct long-time solution. 
This motivates us to add the correction term as introduced in Sec.~\ref{FD-1} and write for the $G^{>}_{xx}$ and $G^{>}_{yx}$ components as, 
\begin{widetext}
\begin{eqnarray}
    &&G_{c,xx}^{>}(t,t')=(-i)^{2}\!\!\int_{t'}^{t}\!\!dt_{1}\!\int_{t_{0}}^{t'}\!\! dt_{2}\sum_{\gamma_{1}\gamma_{2}}\gamma_{1}\gamma_{2}\,\Sigma^{\gamma_{1}\gamma_{2}}(t_{1}\!-\!t_{2})\mathrm{Tr}_S\Big[\sigma_{x}\hat{\mathcal{V}}_0(t\!-\!t_{1})\hat{\sigma}_{x_{\gamma_{1}}}\hat{\mathcal{V}}_0(t_{1}\!-\!t')\sigma_{x}\hat{\mathcal{V}}(t'\!\!-\!t_{2})\hat{\sigma}_{x_{\gamma_{2}}}\rho_{S}(t_{2})\Big],\\
    &&G_{c,yx}^{>}(t,t')=(-i)^{2}\!\!\int_{t'}^{t}\!\!dt_{1}\!\int_{t_{0}}^{t'} \!\!dt_{2}\sum_{\gamma_{1}\gamma_{2}}\gamma_{1}\gamma_{2}\,\Sigma^{\gamma_{1}\gamma_{2}}(t_{1}\!-\!t_{2})\mathrm{Tr}_S\Big[\sigma_{y}\hat{\mathcal{V}}_0(t\!-\!t_{1})\hat{\sigma}_{x_{\gamma_{1}}}\hat{\mathcal{V}}_0(t_{1}\!-\!t')\sigma_{x}\hat{\mathcal{V}}(t'\!\!-\!t_{2})\hat{\sigma}_{x_{\gamma_{2}}}\rho_{S}(t_{2})\Big].
\end{eqnarray}    
\end{widetext}
Recall that these corrections are incorporating the correlations between the spin and the dissipative bath at time $t'$. As a result, the definition of the modified greater Green's function for $t\ge t'$ becomes 
\begin{align}
      \overline{G}^{>}_{xx}(t,t')\!=G^{>}_{xx}(t,t')+G^{>}_{c,xx}(t,t'),\\
      \overline{G}^{>}_{yx}(t,t')\!=G^{>}_{yx}(t,t')+G^{>}_{c,yx}(t,t').
\end{align}
Upon incorporating the correction terms, The EOMs for these modified Green's functions i.e., $\overline{G}^{>}_{xx}(t,t')$ and $\overline{G}^{>}_{yx}(t,t')$ takes the following form, correct up to $O(\lambda^{2})$, 
\begin{align}
    &\frac{\partial}{\partial t}\overline{G}^{>}_{xx}(t,t')=-\omega_{0}\overline{G}^{>}_{yx}(t,t'),\\
    &\frac{\partial}{\partial t}\overline{G}^{>}_{yx}(t,t')=\omega_{0}\overline{G}^{>}_{xx}(t,t')\!-\!2i\!\int_{t_0}^{t}\!\! dt_{1} \Sigma^{K}(t,t_{1})\overline{G}^{>}_{yx}(t_{1},t')\nonumber\\& \,\,\,\,\,\,\,+\!2i\!\!\int_{t_{0}}^{t'}\!\!\!\! dt_{1}\, \Sigma^{>}(t,t_{1})G^{a}_{xy}(t_{1},t')+ O(\lambda^{4}),\label{greater_corr_eom_yx_sb}
\end{align}
where the last equation for $\overline{G}_{yx}^{>}$ takes the typical Kadanoff-Baym type structure with an important inhomogeneous term proportional to $\Sigma^{>} G_{xy}^{a}$ and is missing for the case of Born approximated Green's function. Interestingly in $O(\lambda^2)$, the EOMs for  $\overline{G}^>_{xx}$ and $\overline{G}^>_{yx}$ gets closed. Note that here the $xy$ advanced component i.e., $G^{a}_{xy}(t,t')$ is defined in terms of anti-commutator as (we consider $\xi = -1$ in Eq.~\eqref{retarded_def}), 
\begin{align}
    G^{a}_{xy}(t,t')=i\theta(t'-t)\mathrm{Tr}_S \Big[\sigma_{y}\hat{\mathcal{V}}(t'-t)\Big\{\sigma_{x},\rho(t)\Big\}\Big]. \label{advanced_sb}
\end{align}
Now, similar to CL model, we take the long-time limit in Eq.~\eqref{greater_corr_eom_yx_sb} (recall that in long-time limit all the components of Green's functions are time-translationally invariant) and then perform Fourier transformation with respect to $t-t'$, we receive
\begin{align} 
    &\overline{G}^{>}_{xx}(\omega)=\,\frac{2i\omega_{0}\Sigma^{>}(\omega)G^{a}_{xy}(\omega)}{\Big[\omega^{2}-\omega_{0}^{2}-2\omega \, \Sigma^{K}(\omega)\Big]},\label{greater_longtime_sb_xx}\\ 
    &\overline{G}^{>}_{yx}(\omega)=-\frac{2\omega\Sigma^{>}(\omega)G^{a}_{xy}(\omega)}{\Big[\omega^{2}-\omega_{0}^{2}-2\omega \, \Sigma^{K}(\omega)\Big]}. \label{greater_longtime_sb_yx}
\end{align}
Following similar steps, it is easy to obtain the lesser components from Eq.~\eqref{greater_longtime_sb_xx} and \eqref{greater_longtime_sb_yx} by simply replacing $\Sigma^>(\omega)$ in the numerator with  $\Sigma^<(\omega)$. 
The greater and lesser component of the bath self-energy for the bosonic bath is given by $\Sigma^{>}(\omega)=-i\,J(\omega)\big(1+n_{B}(\omega)\big)$ and $\Sigma^{<}(\omega)=-i\,J(\omega)n_{B}(\omega)$.
Now since $\overline{G}^{>}_{xx}(\omega)$ and $\overline{G}^{>}_{yx}(\omega)$ contains the advanced component $G^{a}_{xy}(\omega)$ in Eq.~\eqref{greater_longtime_sb_xx} and \eqref{greater_longtime_sb_yx}, hence in order to get an explicit form, we need to evaluate the long-time limit of this Green's function $G^{a}_{xy}(t,t')$  as defined in Eq.~\eqref{advanced_sb}. To achieve that, we write down the EOMs of $G^{a}_{xy}(t,t')$ and $G^{a}_{xx}(t,t')$ as they form a closed set of equations. We obtain, 
\begin{align}
    &\frac{\partial}{\partial t'}G^{a}_{xy}(t,t')=\omega_{0}G^{a}_{xx}(t,t')-2i\int_{t_0}^{t'}\!\!\! dt_{1} \Sigma^{K}(t',t_{1})G^{a}_{xy}(t_{1},t),\\
    &\frac{\partial}{\partial t'}G^{a}_{xx}(t,t')=2i\,\delta(t'-t)-\omega_{0}G^{a}_{xy}(t,t').
\end{align}
Now we take the long-time limit in both the EOMs and perform Fourier transformation with respect to $t-t'$ to obtain,
\begin{align}
    G^{a}_{xy}(\omega)=-\frac{2i\omega_{0}}{\Big[\omega^{2}-\omega_{0}^{2}+2\omega\Sigma^{K}(\omega)\Big]}. \label{advanced_longtime_sb}
\end{align}
Interestingly, similar to the CL model, all the advanced and as well as the retarded components for the SB model do not require any correction at order $O(\lambda^{2})$ (see Appendix \ref{Gr_corr_zero} for the proof).
Substituting the expression of $G^{a}_{xy}(\omega)$ from Eq.~\eqref{advanced_longtime_sb} in the long time solution of $G^{>}_{xx}(\omega)$ and $G^{>}_{yx}(\omega)$ in Eq.~\eqref{greater_longtime_sb_xx} and \eqref{greater_longtime_sb_yx}, we receive the final expressions,
\begin{align}
    &\overline{G}^{>}_{xx}(\omega)=\frac{4\omega_{0}^{2}\Sigma^{>}(\omega)}{\Big[(\omega^{2}-\omega_{0}^{2})^{2}-4\omega^{2}\big(\Sigma^{K}(\omega)\big)^{2}\Big]}, \label{greater_xx_sb_longtime}\\ 
    &\overline{G}^{>}_{yx}(\omega)=\frac{4i\omega\omega_0\Sigma^{>}(\omega)}{\Big[(\omega^{2}-\omega_{0}^{2})^{2}-4\omega^{2}\big(\Sigma^{K}(\omega)\big)^{2}\Big]}.\label{greater_yx_sb_longtime}
\end{align}
It is important to note that, the obtained results for the Green's functions in the long-time limit are correct upto $O(\lambda^2)$ (see Appendix \ref{proof_correct_long_time_GF} for the proof). It is easy to observe that Eq.~\eqref{greater_xx_sb_longtime} and the corresponding lesser component satisfy the detailed balance $\overline{G}^{>}_{xx}(\omega)=e^{\beta \omega} \overline{G}^{<}_{xx}(\omega)$ as the self-energy components  i.e., $\Sigma^{>}(\omega), \Sigma^{<}(\omega)$ are also related by the detailed balance relation.  Similarly, the $yx$ component of the greater and lesser Green's function also respects detailed balance condition i.e., $\overline{G}^{>}_{yx}(\omega)=e^{\beta\omega} \overline{G}^{<}_{yx}(\omega)$.


Note that, the final results obtained in Eqs.~\eqref{greater_xx_sb_longtime} and Eq.~\eqref{greater_yx_sb_longtime} exactly match with the work in Ref.~\cite{Agarwalla_2017} obtained using the Majorana fermion approach. In what follows, we extend the above approach to study non-equilibrium steady-state transport where the system can possibly be connected to multiple baths that are subjected to different temperatures. 


\section{Generalisation to non-equilibrium steady-state (NESS) transport-- multi-reservoir scenario} \label{multibath}
In this section, we extend the previously developed approach to the multi-reservoir case where the reservoirs are maintained at different temperatures and assume that a current-carrying non-equilibrium steady state (NESS) sets in \cite{dvira_2,NESS,transport,NESS_2,RevModPhys.94.045006,transport_2}.  We are interested here in computing this NESS current for the CL and the SB models. It can be shown that up to $O(\lambda^{2})$, the effect of multiple reservoirs has an additive effect in the equation of motion for Green's functions. In fact, in the presence of multiple baths, the Dyson Equation for the Born propagator generalizes to, 
\begin{align}
    &\hat{\mathcal{V}}(t-t_0)=\hat{\mathcal{V}}_{0}(t-t_0)-i\int_{t_{0}}^{t} \!\!dt_{1}\!\! \int_{t_{0}}^{t_{1}}\!\! dt_{2}\hat{\mathcal{V}}_{0}(t-t_{1})\nonumber\\&\sum_{\gamma_{1},\gamma_{2};i,j}\!\!\!\gamma_{1}\gamma_{2}\, \Sigma^{\gamma_{1}\gamma_{2}}_{(\rm tot),ij}(t_{1}\!-\!t_{2})\,\hat{S}_{\gamma_{1}}^{i}\hat{\mathcal{V}}_{0}(t_{1}\!-\!t_{2})\hat{S}_{\gamma_{2}}^{j}\hat{\mathcal{V}}(t_{2}\!-\!t_{0}). \label{dyson_V_multibath}
\end{align}
Here, $\Sigma^{\gamma_{1}\gamma_{2}}_{(\rm tot),ij}(t_{1}\!-\!t_{2})\!\!=\!\! \sum_{m} \Sigma^{\gamma_{1}\gamma_{2}}_{(m),ij}(t_{1}\!-\!t_{2})$ where $\Sigma^{\gamma_{1}\gamma_{2}}_{(m),ij}(t_{1}-t_{2})$ is the self energy for the $m$-th reservoir.
As the theory presented here predicts correct results for the correlators up to $O(\lambda^{2})$, by exploiting the additive nature of bath self-energies, we can extend our analysis to calculate the steady-state current for both the CL and the SB model. 

For simplicity, we consider that the system of interest is coupled to two reservoirs (left and right reservoirs). The approach can be easily generalized for multiple reservoir cases. The NESS energy current for arbitrary interacting systems can be calculated following the Meir-Wingreen formula \cite{PhysRevLett.68.2512, Bijay_NEGF,Rammer,haug2008quantum,RevMod_Rammer}, given as 
\begin{equation}
    I_{L}\!\!=\!\!\int_{-\infty}^{+\infty} \frac{d\omega}{4\pi} \, \hbar \omega \, \mathrm{tr}\Big[\xi\,\overline{G}_{S}^{<}(\omega)\Sigma^{>}_{L}(\omega)-\overline{G}_{S}^{>}(\omega)\Sigma^{<}_{L}(\omega)\Big], \label{Meir_win}
\end{equation}
where $I_L$ refers to the current flowing out of the left reservoir and as per our convention current leaving the left reservoir is always considered as positive. The trace in Eq.~\eqref{Meir_win} is over the system's degrees of freedom. Note that the system Green's function $\overline{G}_{S}^{>,<}$ appearing in Eq.~\eqref{Meir_win} corresponds to the two-point correlators of the system operators $S_i, S_j$ (defined in Eq.~\eqref{general_H}) that are coupled to the left bath.  

Let us first discuss the CL model. Recall that, for this case, we already received the exact Green's function $G_{\rm ex}^{>,<}$ for the single bath case. In the presence of the left and right baths, the expression of $\overline{G}^{>}(\omega)$ and $\overline{G}^{<}(\omega)$ become
\begin{align}
    &\overline{G}^{>}(\omega)=G^{r}(\omega)\Big(\Sigma^{>}_{L}(\omega)+\Sigma^{>}_{R}(\omega)\Big)G^{a}(\omega),\\
    &\overline{G}^{<}(\omega)=G^{r}(\omega)\Big(\Sigma^{<}_{L}(\omega)+\Sigma^{<}_{R}(\omega)\Big)G^{a}(\omega).
\end{align}
where the self-energies for the left and right baths appear in an additive manner. Substituting these expressions in the Meir-Wingreen formula in Eq.~\eqref{Meir_win}, we receive the exact Landauer formula for energy current, given as 
\begin{align}
    I_{L}=\int_{-\infty}^{\infty} \frac{d\omega}{4\pi} \, \hbar\omega\, T_{\rm HO}(\omega) \,\Big(n_{L}\big(\omega)-n_{R}(\omega)\Big),
\end{align}
where $T_{\rm HO}(\omega)$ is the transmission function, given as
\begin{eqnarray}
T_{\rm HO}(\omega) &=& G^{r}(\omega)J_{L}(\omega)G^{a}(\omega)J_{R}(\omega),
\end{eqnarray}
where recall that $J_L(\omega), J_R(\omega)$ are now the spectral densities for the left and the right bath, respectively and $J(\omega)$ is defined as $J(\omega)=J_L(\omega)+J_R(\omega)$. 

Next, we calculate the steady-state energy current for the SB model. In this case, in the Meir-Wingreen formula in Eq.~\eqref{Meir_win}, the $G^{>,<}_{xx}$ components enter in place of $G^{>,<}_{S}$. In the presence of two baths, these Green's functions, correct up to $O(\lambda^2)$, are given as
\begin{align}
    &\overline{G}^{>}_{xx}(\omega)=\frac{4\omega_{0}^{2}\,\Big(\Sigma^{>}_{L}(\omega)+\Sigma^{>}_{R}(\omega)\Big)}{\Big[(\omega^{2}-\omega_{0}^{2})^{2}-4 \omega^{2}\big(\Sigma_{L}^{K}(\omega)+\Sigma^{K}_{R}(\omega)\big)^{2}\Big]},\\
    &\overline{G}^{<}_{xx}(\omega)=-\frac{4\omega_{0}^{2}\,\Big(\Sigma_{L}^{<}(\omega)+\Sigma_{R}^{<}(\omega)\Big)}{\Big[(\omega^{2}-\omega_{0}^{2})^{2}-4 \omega^{2}\big(\Sigma_{L}^{K}(\omega)+\Sigma^{K}_{R}(\omega)\big)^{2}\Big]}. 
\end{align}
As a result, we receive the steady-state energy current as \cite{Agarwalla_2017}, 
\begin{equation}
   I_L =\int_{-\infty}^{+\infty} \frac{d\omega}{4\pi}\hspace{0.1cm} \hbar\omega \hspace{0.1cm} {T}_\mathrm{{SB}}(\omega)\big(n_{L}(\omega)-n_{R}(\omega)\big). \label{Landauer-SB}
\end{equation}
where ${T}_\mathrm{{SB}}(\omega)$ is the effective transmission function given as 
\begin{align}
{T}_{\mathrm{SB}}(\omega)=\frac{4\omega_{0}^{2}J_{L}(\omega)J_{R}(\omega)}{\Big[(\omega^{2}-\omega_{0}^{2})^{2}-4 \omega^{2}\big(\Sigma_{L}^{K}(\omega)+\Sigma^{K}_{R}(\omega)\big)^{2}\Big]}.
\end{align}
This form of current in Eq.~\eqref{Landauer-SB} is similar to the exact Landauer formula with a crucial difference of $T_{\mathrm{SB}}(\omega)$ being temperature dependent via the Keldysh self-energy component which is a typical signature for an interacting model. 


\section{Summary} \label{summary}
In this work, we show that the definition of non-equilibrium Green's function based on the Born approximation, in the context of the open quantum systems, does not provide correct results to the leading order $O(\lambda^2)$ in the system bath coupling. This is due to the fact that the Born approximated definition misses a crucial correlation between the system and the bath. 
We have explicitly identified this problem and proposed a correction term to the standard Born definition to get a consistent theory that is correct up to $O(\lambda^2)$. Such a correction term precisely captures the exchange of energy quanta between the system and the bath. More explicitly, for both the Caldeira-Leggett and the spin-boson model, we show that the correction terms lead to a standard Kadanoff-Baym type equation for the greater/lesser Green's function which correctly produces the long-time limit and preserves the detailed balance condition in thermal equilibrium. We have further extended this approach to the non-equilibrium multi-terminal scenario and calculated the steady state energy current following the Meir-Weingreen formula. Interestingly, we obtain the exact Landauer formula for the Calderia-Leggett model, whereas, for the spin-boson model, we obtain an effective Landauer-like formula in the long-time limit. Future work will be directed toward understanding the impact of system-bath coupling terms beyond the leading order of the system-bath coupling on the Green's function. A useful and practical way to analyze this would be either following the recently proposed non-crossing approximated approach for open quantum systems \cite{Marco_schiro} or the reaction-coordinate approach \cite{Nazir_2018}. 



\section*{Acknowledgements}  
The authors would like to acknowledge Sakil Khan, Sandipan Mohanta, Archak Purkayastha, and Madhumita Saha for many fruitful discussions. BKA acknowledges the MATRICS grant MTR/2020/000472 from SERB, Government of India. BKA also thanks the Shastri Indo-Canadian Institute for providing financial support for this research work in the form of a Shastri Institutional Collaborative Research Grant (SICRG). KG would like to acknowledge the University Grants Commission (UGC) of India for her research fellowship (Ref. No.- 211610104535).  The authors would like to thank the International Centre for Theoretical Sciences (ICTS) for organizing the program - Periodically and quasi-periodically driven complex systems (code: ICTS/pdcs2023/6) where many interesting discussions related to this project took place. 
\begin{widetext}
\appendix 
\section{Derivation for the standard definition of two-point correlator under the Born approximation} \label{standard_def_proof}
The general expression of a two-point and two-time correlation function for two operators $A$ and $B$ is given by the following,
\begin{align}
    &\langle A(t)B(t') \rangle \\
    & = \mathrm{Tr}_{S,R}\Big[A(t)B(t')\rho_{\rm tot}(t_{0})\Big]\\
    &=\mathrm{Tr}_{S,R}\Big[e^{iHt}Ae^{-iHt}e^{iHt'}Be^{-iHt'}\rho_{\rm tot}(t_{0})\Big]\\
    &=\theta(t\!-\!t')\mathrm{Tr}_{S,R}\Big[Ae^{-iH(t-t')}Be^{-iHt'}\rho_{\rm tot}(t_{0})e^{iHt'}e^{iH(t-t')}\Big]+\theta(t'\!-\!t)\mathrm{Tr}_{S,R}\Big[Be^{-iH(t'-t)}e^{-iHt}\rho_{\rm tot}(t_{0})e^{iHt}e^{iH(t'\!-t)}\Big]\\
    &=\theta(t-t')\mathrm{Tr}_{S,R}\Big[Ae^{-iH(t-t')}B\rho_{\rm tot}(t')e^{iH(t-t')}\Big]+\theta(t'-t)\mathrm{Tr}_{S,R}\Big[Be^{-iH(t'\!\!-\!t)}\rho_{\rm tot}(t')Ae^{iH(t-t')}\Big].
\end{align}
So far the above expressions are exact. Now let us consider making the Born approximation i.e., $\rho_{\rm tot}(t_{m})=\rho_{S}(t_m)\otimes \bar{\rho}_{R}$ ($t_m$ represents the minimum of  $t$ or $t'$) in the above expression and expand the exponential order by order and finally trace out the reservoir. Then the final expression of the two-point correlator under Born approximation reduces to,
\begin{align}
    \langle A(t)B(t') \rangle =\theta(t-t')\mathrm{Tr}_{S}\Big[A\hat{\mathcal{V}}(t-t')\big(B\rho_{S}(t')\big)\Big]+\theta(t'-t)\mathrm{Tr}_{S}\Big[B\hat{\mathcal{V}}(t'-t)\big(\rho_{S}(t)A\big)\Big]=iG^{>}_{AB}(t,t').
\end{align}
Similarly, one can show that
\begin{align}
    \langle B(t')A(t) \rangle =\theta(t-t')\mathrm{Tr}_{S}\Big[A\hat{\mathcal{V}}(t-t')\big(\rho_{S}(t')B\big)\Big]+\theta(t'-t)\mathrm{Tr}_{S}\Big[B\hat{\mathcal{V}}(t'-t)\big(A\rho_{S}(t)\big)\Big]=-i\xi G^{<}_{AB}(t,t'),
\end{align}
where $\xi= \pm 1$, as discussed in the main text. 
\section{Proof for the modified definition for the greater Green's function that is correct up to $O(\lambda^{2})$} \label{Dyson_eqn_G_prrof_NEGF}
In the main text, we have obtained the modified definition of the greater Green's function in Eq.~\eqref{modified_correlator}. In this appendix, we will show that the modified definition is exactly correct up to $O(\lambda^{2})$. 
First, we start with the definition of the modified Green's function $G^{>}_{AB}(t,t')$ for $t\ge t'$ and write the Dyson equation as follows,
\begin{align}
    G^{>}_{AB}(t,t') \!&=\! -i\mathrm{Tr}_{S}\big\{A\hat{\mathcal{V}}(t\!-\!t')B\rho_{S}(t')\!\big\} \!-i \langle A(t) B(t') \rangle_{c}\\
    &=-i\mathrm{Tr}_{S}\Big[A\hat{\mathcal{V}}_{0}(t-t')\big(B\hat{\mathcal{V}}_{0}(t')\rho_{S}(t_{0})\big)\Big]
    \!\!\nonumber\\&+(-i)^{2}\!\!\int_{t_{0}}^{t'}\!\!\!\! dt_{1}\! \int_{t_{0}}^{t_{1}}\!\!\!\! dt_{2}\!\!\!\sum_{\gamma_{1},\gamma_{2};i,j}\!\! \gamma_{1}\gamma_{2}\,\Sigma^{\gamma_{1}\gamma_{2}}_{ij}(t_{1}\!-\!t_{2})\,\mathrm{Tr_{S}}\Big[A\,\hat{\mathcal{V}}_{0}(t\!-\!t')B\hat{\mathcal{V}}_{0}(t'\!-\!t_{1})\hat{S}^{i}_{\gamma_{1}}\hat{\mathcal{V}}_{0}(t_{1}\!-\!t_{2})\hat{S}^{j}_{\gamma_{2}}\rho_{S}(t_{2})\Big]\nonumber\\&+(-i)^{2}\!\int_{t'}^{t}\!\! dt_{1}\! \int_{t'}^{t_{1}}\!\!\!\! dt_{2}\!\!\sum_{\gamma_{1},\gamma_{2};i,j} \!\!\!\gamma_{1}\gamma_{2}\,\Sigma^{\gamma_{1}\gamma_{2}}_{ij}(t_{1}\!\!-\!\!t_{2})\,\mathrm{Tr_{S}}\Big[A\hat{\mathcal{V}}_{0}(t\!-\!t_{1})\hat{S}^{i}_{\gamma_{1}}\hat{\mathcal{V}}_{0}(t_{1}\!-\!t_{2})\hat{S}^{j}_{\gamma_{2}}\hat{\mathcal{V}}(t_{2}\!-\!t')B\rho_{S}(t')\Big]\nonumber\\&+(-i)^{2}\!\!\int_{t'}^{t} \!\!dt_{1}\! \int_{t_{0}}^{t'}\!\!\!\! dt_{2}\!\!\!\sum_{\gamma_{1},\gamma_{2};i,j} \!\!\!\gamma_{1}\gamma_{2}\,\Sigma^{\gamma_{1}\gamma_{2}}_{ij}(t_{1}\!-\!t_{2})\,\mathrm{\mathrm{Tr_{S}}}\Big[A\hat{\mathcal{V}}_{0}(t\!-\!t_{1})\hat{S}^{i}_{\gamma_{1}}\hat{\mathcal{V}}_{0}(t_{1}\!-\!t')B\hat{\mathcal{V}}(t'\!-\!t_{2})\hat{S}^{j}_{\gamma_{2}}\rho_{S}(t_{2})\Big].\label{dyson_NEGF_G_AB}
\end{align}
Here the first term in Eq.~\eqref{dyson_NEGF_G_AB} is the bare Green's function and all the other three terms are order $O(\lambda^{2})$ contribution to the greater Green's function. The first, second, and third terms in  
Eq.~\eqref{dyson_NEGF_G_AB} are obtained in Eq.~\eqref{diagram_expansion} from the Born approximated definition in Eq.~\eqref{greater_def}. The last term in Eq.~\eqref{dyson_NEGF_G_AB} is our correction term obtained in the main text, as given in Eq.~\eqref{general_cor}.
In all the terms we  replace the Born propagator $\hat{\mathcal{V}}$ in Eq~\eqref{dyson_NEGF_G_AB} by the free propagator $\hat{\mathcal{V}}_0$ to consider terms only up to order $O(\lambda^{2})$. We therefore receive,
\begin{align}
    G^{>}_{AB}(t,t')&=-i\mathrm{Tr}_{S}\Big[A\hat{\mathcal{V}}_{0}(t-t')\big(B\hat{\mathcal{V}}_{0}(t')\rho_{S}(t_{0})\big)\Big]
    \!\!\nonumber\\&+(-i)^{2}\!\!\int_{t_{0}}^{t'}\!\!\!\! dt_{1}\! \int_{t_{0}}^{t_{1}}\!\!\!\! dt_{2}\!\!\!\sum_{\gamma_{1},\gamma_{2};i,j}\!\! \gamma_{1}\gamma_{2}\,\Sigma^{\gamma_{1}\gamma_{2}}_{ij}(t_{1}\!-\!t_{2})\,\mathrm{Tr_{S}}\Big[A\,\hat{\mathcal{V}}_{0}(t\!-\!t')B\hat{\mathcal{V}}_{0}(t'\!-\!t_{1})\hat{S}^{i}_{\gamma_{1}}\hat{\mathcal{V}}_{0}(t_{1}\!-\!t_{2})\hat{S}^{j}_{\gamma_{2}}\hat{\mathcal{V}}_0(t_2-t_0)\rho_{S}(t_{0})\Big]\nonumber\\&+(-i)^{2}\!\int_{t'}^{t}\!\! dt_{1}\! \int_{t'}^{t_{1}}\!\!\!\! dt_{2}\!\!\sum_{\gamma_{1},\gamma_{2};i,j} \!\!\!\gamma_{1}\gamma_{2}\,\Sigma^{\gamma_{1}\gamma_{2}}_{ij}(t_{1}\!\!-\!\!t_{2})\,\mathrm{Tr_{S}}\Big[A\,\hat{\mathcal{V}}_{0}(t\!-\!t_{1})\hat{S}^{i}_{\gamma_{1}}\hat{\mathcal{V}}_{0}(t_{1}\!-\!t_{2})\hat{S}^{j}_{\gamma_{2}}\hat{\mathcal{V}}_0(t_{2}\!-\!t')B\hat{\mathcal{V}}_0(t'\!-\!t_0)\rho_{S}(t_0)\Big]\nonumber\\&+(-i)^{2}\!\!\int_{t'}^{t} \!\!dt_{1}\! \int_{t_{0}}^{t'}\!\!\!\! dt_{2}\!\!\!\sum_{\gamma_{1},\gamma_{2};i,j} \!\!\!\gamma_{1}\gamma_{2}\,\Sigma^{\gamma_{1}\gamma_{2}}_{ij}(t_{1}\!-\!t_{2})\,\mathrm{\mathrm{Tr_{S}}}\Big[A\,\hat{\mathcal{V}}_{0}(t\!-\!t_{1})\hat{S}^{i}_{\gamma_{1}}\hat{\mathcal{V}}_{0}(t_{1}\!-\!t')B\hat{\mathcal{V}}_0(t'\!-\!t_{2})\hat{S}^{j}_{\gamma_{2}}\hat{\mathcal{V}}_{0}(t_2\!-\!t_0)\rho_{S}(t_{0})\Big].\label{NEGF_expr}
\end{align}
Substituting the expression of $\hat{\mathcal{V}}_0(t)[\bullet]=e^{-iH_{S}t}\bullet e^{iH_St}$ in Eq.~\eqref{NEGF_expr}, we simplify the terms using the interaction picture representation of the operators. Hence we obtain,
\begin{align}
    G^{>}_{AB}(t,t')&=-i\langle A_{I}(t)B_{I}(t') \rangle_{0}+(-i)^{2}\!\int_{t_{0}}^{t'}\!\!\! dt_{1} \int_{t_{0}}^{t_{1}} \!\!\!dt_{2}\!\!\!\sum_{\gamma_{1},\gamma_{2};i,j}\!\!\!\! \gamma_{1}\gamma_{2}\,\Sigma^{\gamma_{1}\gamma_{2}}_{ij}(t_{1}\!-\!t_{2})\mathrm{Tr_{S}}\Big[A_{I}(t)B_{I}(t')\hat{S}^{i}_{I,\gamma_{1}}(t_{1})\hat{S}^{j}_{I,\gamma_{2}}(t_{2})\rho_{S}(t_{0})\Big]\nonumber\\&+(-i)^{2}\int_{t'}^{t} dt_{1} \int_{t'}^{t_{1}} dt_{2}\sum_{\gamma_{1},\gamma_{2};i,j} \gamma_{1}\gamma_{2}\Sigma^{\gamma_{1}\gamma_{2}}_{ij}(t_{1}-t_{2})\mathrm{Tr_{S}}\Big[A_{I}(t)\hat{S}^{i}_{I,\gamma_{1}}(t_{1})\hat{S}^{j}_{I,\gamma_{2}}(t_{2})B_{I}(t')\rho_{S}(t_{0})\Big]\nonumber\\&+(-i)^{2}\int_{t'}^{t} dt_{1} \int_{t_{0}}^{t'} dt_{2}\sum_{\gamma_{1},\gamma_{2};i,j} \gamma_{1}\gamma_{2}\Sigma^{\gamma_{1}\gamma_{2}}_{ij}(t_{1}-t_{2})\mathrm{\mathrm{Tr_{S}}}\Big[A_{I}(t)\hat{S}^{i}_{I,\gamma_{1}}(t_{1})B_{I}(t')\hat{S}^{j}_{I,\gamma_{2}}(t_{2})\rho_{S}(t_{0})\Big].\label{time_ordering_GAB}
\end{align}
The different integration limits in the last three terms in Eq.~\eqref{time_ordering_GAB} leads to different time orderings for $t_1,\,t_2$, which is represented in Fig.~\ref{time_ordering}. One can easily observe that the second, third, and fourth term in Eq.~\eqref{time_ordering_GAB} corresponds to region-I, II, and III respectively.
\begin{figure}
\centering
\begin{minipage}{0.4\textwidth}
   \centering  \includegraphics[width=2.5in,height=2.5in]{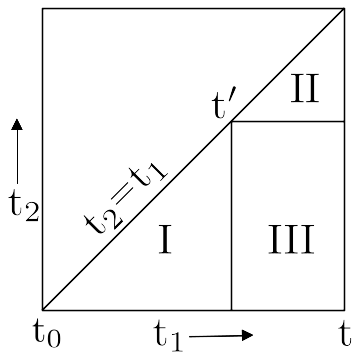}
  \captionof{figure}{Different time orderings of $t_{1}$ and $t_{2}$, represented in Eq.~\eqref{time_ordering_GAB}: Region (I) and (II) reflect the Born approximated regime whereas the region (III) is the correction added to the Born approximated definition of the Green's function.}
  \label{time_ordering}
\end{minipage}%
\hspace{1.8cm}
\begin{minipage}{0.4\textwidth}
   \centering
  \includegraphics[width=2.55in,height=1.34in]{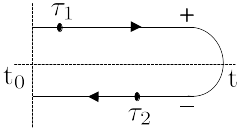}
  \captionof{figure}{Keldysh contour: The upper (lower) branch of the contour  corresponds to the forward (backward) time evolution and denoted by + (-) sign for the indices $\gamma_i$. Here $\tau_1$ and $\tau_2$ are the contour times corresponding to the real-time $t_1$ and $t_2$ respectively. The dashed horizontal line in the middle of the branches is the real-time axis starting from the initial time $t_0$, ending with the maximum time $t$.}\label{Keldysh_contour}
\end{minipage}
\end{figure}

From Fig.~\ref{time_ordering}, one can see that the three regions (I+II+III) form a triangular area where $t_1$ ranges from initial time $t_0$ to the maximum time $t$ and $t_2$ ranges from initial time $t_0$ to the time $t_1$. Hence we can club the three integrations in Eq.~\eqref{time_ordering_GAB} by using the time ordering operator $T_R$ as the following, 
\begin{align}
     G^{>}_{AB}(t,t')&=-i\langle A_{I}(t)B_{I}(t') \rangle_{0}+(-i)^{2}\!\int_{t_{0}}^{t} \!\!\!dt_{1} \!\!\int_{t_{0}}^{t_{1}}\!\!\! dt_{2}\!\!\!\sum_{\gamma_{1},\gamma_{2};i,j} \!\!\!\!\gamma_{1}\gamma_{2}\,\Sigma^{\gamma_{1}\gamma_{2}}_{ij}(t_{1}\!-\!t_{2})\mathrm{Tr_{S}}\Big[T_{R}A_{I}(t)B_{I}(t')\hat{S}^{i}_{I,\gamma_{1}}(t_{1})\hat{S}^{j}_{I,\gamma_{2}}(t_{2})\rho_{S}(t_{0})\Big]\nonumber\\
    &=-i\langle A_{I}(t)B_{I}(t') \rangle_{0}+\frac{(-i)^{2}}{2}\!\int_{t_{0}}^{t} \!\!\!dt_{1}\!\!\! \int_{t_{0}}^{t} dt_{2}\!\!\!\!\sum_{\gamma_{1},\gamma_{2};i,j}\!\!\! \gamma_{1}\gamma_{2}\,\Sigma^{\gamma_{1}\gamma_{2}}_{ij}(t_{1}\!-\!t_{2})\mathrm{Tr_{S}}\Big[T_{R}A_{I}(t)B_{I}(t')\hat{S}^{i}_{I,\gamma_{1}}(t_{1})\hat{S}^{j}_{I,\gamma_{2}}(t_{2})\rho_{S}(t_{0})\Big].
\end{align}
It is now possible to map the real-time $t,\,t'$ to the Keldysh contour time variable $\tau,\,\tau'$. By considering $\gamma_i = \pm 1$, ($i=1,2$) as the two different branches of the Keldysh contour shown in Fig.~\ref{Keldysh_contour}, we can write 
\begin{align}
   G_{AB}(\tau,\tau') = -i\langle A_{I}(\tau)B_{I}(\tau') \rangle_{0}+\frac{(-i)^{2}}{2!}\sum_{i,j}\int_{C}\!\!\! d\tau_{1}\!\!\int_{C}\!\!\!d\tau_{2}\,\Sigma(\tau_{1},\tau_{2})\,\mathrm{Tr}_{S}\Big[T_{C}A_{I}(\tau)B_{I}(\tau')S_{I}^{i}(\tau_{1}) S_{I}^{j}(\tau_{2}) \rho_{S}(t_{0})\Big],\label{G_AB}
\end{align}
where $T_C$ is the Keldysh contour time-ordering operator which orders the operator according to their position on the Keldysh contour. Operators appearing later on the contour are placed to the left. It is now easy to see that, the above expression in Eq.~\eqref{G_AB} corresponds to the $O(\lambda^0)$ (bare Green's function) and the $O(\lambda^{2})$ contribution to the contour ordered Green's function $\tilde{G}_{AB}(\tau,\tau')$, where $\tilde{G}_{AB}(\tau,\tau')$ is defined as,
\begin{align}
    \tilde{G}_{AB}(\tau,\tau')=-i\mathrm{Tr}_{S,R}\Big[T_{C}A_{I}(\tau)B_{I}(\tau')e^{-i\sum_{i,j} \int_{C} d\tau_{1}\lambda  \sum_{i} g_{i} S_{I}^{i}(\tau_{1})\otimes R_{I}^{i}(\tau_{1})}\rho(t_{0})\Big].\label{contour
    -ordered_GF}
\end{align}
In summary, starting from the modified definition of the greater Green's function $G^{>}_{AB}(t,t')$, we finally obtained the contour order Green's function $\tilde{G}_{AB}(\tau,\tau')$ up to $O(\lambda^{2})$. Hence it is proved that, after adding the correction term introduced in Eq.~\eqref{general_cor}, to the Born approximated standard definition of the greater Green's function in Eq~\eqref{greater_def}, the modified greater Green's function is correct up to $O(\lambda^{2})$.
\section{General proof to show that $O(\lambda^{2})$ correct equation of motion of Green's function results $O(\lambda^{2})$ correct long-time solution} \label{proof_correct_long_time_GF}
In this appendix, we will show that in the context of Non-equilibrium Green's function, an equation of motion (EOM) that is correct up to $O(\lambda^{2})$ is sufficient for obtaining steady-state Green's function correct up to the same order i.e., $O(\lambda^{2})$. Below we provide the proof. 

First we write $G^{>}(t,t')$ and $\frac{\partial }{\partial t}G^{>}(t,t')$ as,
\begin{align}
    &G^{>}(t,t')=\sum_{m=1}^{\infty} \lambda^{2m} \, G^{>}_{(2m)}(t,t'), \label{GF_Expand}\\
    &\frac{\partial}{\partial t} G^{>}(t,t')=\hat{\mathcal{L}}(t,t')\Big[G^{>}(t,t')\Big]+\lambda^{2}\mathcal{F}(t,t')=\sum_{m=1}^{\infty} \lambda^{2m} \, \hat{\mathcal{L}}_{(2m)}(t,t')\Big[G^{>}(t,t')\Big]+\lambda^{2}\hat{\mathcal{F}}(t,t'). \label{EOM_GF_Expand}
\end{align}
Here $G^{>}_{(2m)}(t,t')$ is the $O(\lambda^{2m})$ contribution of $G^{>}(t,t')$. Also $\hat{\mathcal{L}}(t,t')[\bullet]=\sum_{m=1}^{\infty} \lambda^{2m} \, \hat{\mathcal{L}}_{(2m)}(t,t')[\bullet]$ is a superoperator expanded order by order in $\lambda$. It contains the homogeneous terms in the above differential equation in Eq.~\eqref{EOM_GF_Expand}. $\lambda^{2}\mathcal{F}(t,t')$ corresponds to the inhomogeneous terms in eq.~\eqref{EOM_GF_Expand} and it is already of $O(\lambda^{2})$. Note that, for simplicity, we consider here a first-order differential equation for the Green's function. However, our proof can be easily generalized to higher-order differential equations as well. Now substituting $G^{>}(t,t')$ from Eq.~\eqref{GF_Expand} in the Eq.~\eqref{EOM_GF_Expand}, we receive,
\begin{align}
    \frac{\partial}{\partial t} \sum_{p=1}^{\infty} \lambda^{2p} \, G^{>}_{(2p)}(t,t')=\sum_{m=1}^{\infty} \lambda^{2m} \, \hat{\mathcal{L}}_{(2m)}(t,t')\Bigg[\sum_{n=1}^{\infty} \lambda^{2n} \, G^{>}_{(2n)}(t,t')\Bigg]+\lambda^{2}\mathcal{F}(t,t'). \label{EOM_GF_Expanded_expression}
\end{align}
Next we take the long-time limit $(t,t'\rightarrow \infty$ but $t-t'$ is finite) in Eq.~\eqref{EOM_GF_Expanded_expression}. In the steady state, Green's function $G^{>}(t,t')$ and the inhomogeneous term $\mathcal{F}(t,t')$  will have time-translational invariance i.e., $\big(G^{>,<}(t,t')=G^{>,<}(t-t')\big)$ and $\mathcal{F}(t,t')=\mathcal{F}(t-t')$. The superoperator $\hat{\mathcal{L}}(t,t')[\bullet]$ also has a long time limit $\hat{\mathcal{L}}^{SS}[\bullet]=\sum_{m=1}^{\infty} \lambda^{2m} \, \hat{\mathcal{L}}^{SS}_{(2m)}[\bullet]$. We then receive, 
\begin{align}
    \frac{\partial}{\partial t} \sum_{p=1}^{\infty} \lambda^{2p} \, G^{>}_{(2p)}(t-t')=\sum_{m=1}^{\infty} \lambda^{2m} \, \hat{\mathcal{L}}_{(2m)}^{SS}\Bigg[\sum_{n=1}^{\infty} \lambda^{2n} \, G^{>}_{(2n)}(t-t')\Bigg]+\lambda^{2}\mathcal{F}(t-t'). \label{SS_EOM_Expand}
\end{align}
Performing the Fourier transformation with respect to $t-t'=\tau$ in Eq.~\eqref{SS_EOM_Expand}, we obtain 
\begin{align}
    -i\omega \sum_{p=1}^{\infty} \lambda^{2p} G^{>}_{(2p)}(\omega)= \sum_{m=1}^{\infty} \lambda^{2m} \, \hat{\Tilde{\mathcal{L}}}_{(2m)}^{SS}\Bigg[\sum_{n=1}^{\infty} \lambda^{2n} \, G^{>}_{(2n)}(\omega)\Bigg]+\lambda^{2}\tilde{\mathcal{F}}(\omega). \label{FT_EOM_Expanded}
\end{align}
Here $\hat{\Tilde{\mathcal{L}}}^{SS}[\bullet]$ and $\tilde{\mathcal{F}}(\omega)$ represent the superoperators after performing the Fourier transform of $\hat{\mathcal{L}}^{SS}[\bullet]$. Now equating the terms of $\lambda^{(0)}$ and $\lambda^{(2)}$ in Eq.~\eqref{FT_EOM_Expanded} respectively,
\begin{align}
    &-i\omega \, G^{>}_{(0)}(\omega)=\hat{\Tilde{\mathcal{L}}}_{(0)}^{SS}\big[G^{>}_{(0)}(\omega)\big],\\
    &-i\omega \, \lambda^{2} \, G^{>}_{(2)}(\omega)= \lambda^{2} \, \hat{\Tilde{\mathcal{L}}}^{SS}_{(2)}\big[G^{>}_{(0)}(\omega)\big]+\lambda^{2} \, \hat{\Tilde{\mathcal{L}}}^{SS}_{0}\big[G^{>}_{(2)}(\omega)\big]+\lambda^{2}\tilde{\mathcal{F}}(\omega). \label{lambda2_order}
\end{align}
Now adding the above two equations, we obtain,
\begin{align}
    -i\omega \, \Big(G^{>}_{(0)}(\omega)+\lambda^{2}\, G^{>}_{(2)}(\omega)\Big)=\hat{\Tilde{\mathcal{L}}}_{(0)}^{SS}\big[G^{>}_{(0)}(\omega)+\lambda^{2}\, G^{>}_{(2)}(\omega)\big]+\lambda^{2} \, \hat{\Tilde{\mathcal{L}}}^{SS}_{(2)}\big[G^{>}_{(0)}(\omega)\big]+\lambda^{2}\tilde{\mathcal{F}}(\omega). \label{final_EOM_GF}
\end{align}
Let's define $G^{>}(\omega)=G^{>}_{(0)}(\omega)+\lambda^{2}\, G^{>}_{(2)}(\omega)$, where $G^{>}(\omega)$ is the correct greater Green's function up to $O(\lambda^{2})$. Our aim is to obtain the $G^{>}(\omega)$. Hence we will replace the $G^{>}_{(0)}(\omega)$ by $G^{>}(\omega)$ in the second term in the RHS of Eq.~\eqref{final_EOM_GF} as the second term is already of order $O(\lambda^2)$. So finally we receive,
\begin{align}   
    -i\omega \, G^{>}(\omega)=\hat{\Tilde{\mathcal{L}}}_{(0)}^{SS}[G^{>}(\omega)]+\lambda^{2} \, \hat{\Tilde{\mathcal{L}}}^{SS}_{(2)}\big[G^{>}(\omega)\big]+\lambda^{2}\tilde{\mathcal{F}}(\omega). \label{correct_G_order_lambda}
\end{align}
From Eq.~\eqref{correct_G_order_lambda}, one can obtain the steady state solution of the greater component of Green's function $G^{>}(\omega)$ which is correct up to $O(\lambda^{2})$. For obtaining that we require only $\hat{\Tilde{\mathcal{L}}}_{(0)}^{SS}$, $\hat{\Tilde{\mathcal{L}}}_{(2)}^{SS}$ and $\mathcal{F}$ and all of these are of order $O(\lambda^{2})$. Hence it is proved that to obtain correct steady-state Green's functions up to order $O(\lambda^{2})$, an EOM correct up to $O(\lambda^{2})$ is sufficient. Needless to mention, this proof trivially goes through for other components of Green's functions. 
\section{Exact Dyson equation for the retarded Green's function $G^{r}(t,t')$ for the Caldeira-Leggett model}\label{exact_dyson_Gr}
In this appendix, we show that for the Caldeira-Leggett (CL) model, starting from the Born approximated definition for the retarded Green's function, interestingly, we obtain the exact Dyson equation, correct up to all orders of the system-bath coupling.
We start with the definition for the Born approximated retarded correlator, as defined in Eq.~\eqref{retarded_CL_def},
\begin{align}
    G^{r}(t,t')= -i\,\theta(t-t')\mathrm{Tr}_{S}\Big[x\hat{\mathcal{V}}(t-t')\big[x,\rho_{S}(t')\big]\Big].
    \label{Gr-app}
\end{align}
Now, substituting the Dyson equation of $\hat{\mathcal{V}}(t-t')$ as defined in Eq.~\eqref{dyson_V} in Eq.~\eqref{Gr-app}, we receive,
\begin{align}
    G^{r}(t,t')&= -i\,\theta(t\!-\!t')\mathrm{Tr}_{S}\Big[x\hat{\mathcal{V}}_{0}(t\!-\!t')\big[x,\hat{\mathcal{V}}_{0}(t'\!-\!t_0)\rho_{S}(t_0)\big]\Big]\nonumber\\&+\!(-i)^{2}\theta(t-t')\!\int_{t'}^{t}\!\!\! dt_{1} \!\!\int_{t'}^{t_{1}}\!\!\! dt_{2}\!\!\sum_{\gamma_{1},\gamma_{2}}\!\!\!\gamma_{1}\gamma_{2}\,\Sigma^{\gamma_{1}\gamma_{2}}(t_{1}\!-\!t_{2})\mathrm{Tr}_{S}\Big[x\,\hat{\mathcal{V}}_{0}(t\!-\!t_{1}) \hat{x}_{\gamma_{1}}\,\hat{\mathcal{V}}_{0}(t_{1}\!-\!t_{2})\hat{x}_{\gamma_{2}}\,\hat{\mathcal{V}}(t_{2}\!-\!t')\big[x,\rho_{S}(t')\big]\Big]\nonumber\\&+\!\!(-i)^{2}\theta(t\!-\!t')\!\!\int_{t_{0}}^{t'} \!\!\!\!dt_{1}\!\! \int_{t_{0}}^{t_{1}}\!\!\!\! dt_{2}\!\!\sum_{\gamma_{1},\gamma_{2}}\!\!\!\gamma_{1}\gamma_{2}\,\Sigma^{\gamma_{1}\gamma_{2}}(t_{1}\!-\!t_{2})\mathrm{Tr}_{S}\Big[x\,\hat{\mathcal{V}}_{0}(t\!-\!t')\big[x,\hat{\mathcal{V}}_0(t'\!-\!t_1)\hat{x}_{\gamma_{1}}\hat{\mathcal{V}}_{0}(t_{1}\!-\!t_{2})\hat{x}_{\gamma_{2}}\hat{\mathcal{V}}(t_{2}\!-\!t')\rho_{S}(t_0)\big]\Big]. \label{Gr_dyson}
\end{align}
The first term is the bare retarded correlator, denoted as $G^{r}_{0}(t,t')$. Now, we will simplify the second and the third terms. Let's start with the second term and express the bare propagator $\hat{\mathcal{V}}_0$ using their original form as $\hat{\mathcal{V}}_0(t)\bullet=e^{-iHt}\bullet e^{iHt}$, we get
\begin{align}
    &(-i)^{2}\theta(t-t')\int_{t'}^{t} dt_{1} \int_{t'}^{t_{1}} dt_{2}\sum_{\gamma_{1},\gamma_{2}}\gamma_{1}\gamma_{2}\Sigma^{\gamma_{1}\gamma_{2}}(t_{1}-t_{2})\mathrm{Tr}_{S}\Big[x\hat{\mathcal{V}}_{0}(t-t_{1}) \, \,\hat{x}_{\gamma_{1}}\,\hat{\mathcal{V}}_{0}(t_{1}-t_{2})\,\hat{x}_{\gamma_{2}}\,\hat{\mathcal{V}}(t_{2}-t')\big[x,\rho_{S}(t')\big]\Big]\\
    &=(-i)^{2}\theta(t-t')\int_{t'}^{t} dt_{1} \int_{t'}^{t_{1}} dt_{2}\sum_{\gamma_{1},\gamma_{2}}\gamma_{1}\gamma_{2}\Sigma^{\gamma_{1}\gamma_{2}}(t_{1}-t_{2})\mathrm{Tr}_{S}\Big[x_{I}(t-t_1) \, \,\hat{x}_{\gamma_{1}}\,\hat{\mathcal{V}}_{0}(t_{1}-t_{2})\,\hat{x}_{\gamma_{2}}\,\hat{\mathcal{V}}(t_{2}-t')\big[x,\rho_{S}(t')\big]\Big]. \label{second_term_step_2}
\end{align}
Here the subscript $I$ denotes the interaction picture representation of the operator $x$. We can relate the interaction picture representation of $x$ to the Schr$\mathrm{\ddot{o}}$dinger picture by the following,
\begin{align}
    x_{I}(t-t_1)=p \,G^{r}_{0}(t-t_1)\,+ x \, \frac{\partial }{\partial t}G^{r}_{0}(t-t_1)=p \,G^{r}_{0}(t-t_1)\,+ x \, F^{r}_{0}(t-t_1),
    \label{XI_CL}
\end{align}
where $p= \dot{x}$ is the momentum operator and recall that we have considered $M=1$. Substituting this expression of $x_{I}(t-t_1)$ in Eq.~\eqref{second_term_step_2}, we get
\begin{align}
    &=(-i)^{2}\theta(t\!-\!t')\int_{t'}^{t}\!\!\! dt_{1}\!\! \int_{t'}^{t_{1}} \!\!\!\!dt_{2}\,G_{0}^{r}(t-t_1)\sum_{\gamma_{1},\gamma_{2}}\!\!\gamma_{1}\gamma_{2}\,\Sigma^{\gamma_{1}\gamma_{2}}(t_{1}\!-\!t_{2})\mathrm{Tr}_{S}\Big[p \, \,\hat{x}_{\gamma_{1}}\,\hat{\mathcal{V}}_{0}(t_{1}\!-\!t_{2})\,\hat{x}_{\gamma_{2}}\,\hat{\mathcal{V}}(t_{2}\!-\!t')\big[x,\rho_{S}(t')\big]\Big]\nonumber\\& +(-i)^{2}\theta(t\!-\!t')\!\int_{t'}^{t} \!\!\!dt_{1}\!\! \int_{t'}^{t_{1}}\!\!\!\! dt_{2}\,F^{r}_{0}(t-t_1)\,\sum_{\gamma_{1},\gamma_{2}}\!\!\!\gamma_{1}\gamma_{2}\,\Sigma^{\gamma_{1}\gamma_{2}}(t_{1}\!-\!t_{2})\mathrm{Tr}_{S}\Big[x\,\hat{x}_{\gamma_{1}}\,\hat{\mathcal{V}}_{0}(t_{1}\!-\!t_{2})\,\hat{x}_{\gamma_{2}}\,\hat{\mathcal{V}}(t_{2}\!-\!t')\big[x,\rho_{S}(t')\big]\Big]. 
\end{align}
Now let us perform the summation over $\gamma_1$ first. We obtain,
\begin{align}
     &=(-i)^{2}\theta(t-t')\int_{t'}^{t} dt_{1} \int_{t'}^{t_{1}} dt_{2}\,G_{0}^{r}(t-t_1)\sum_{\gamma_{2}}\gamma_{2}\Sigma^{+\gamma_{2}}(t_{1}-t_{2})\mathrm{Tr}_{S}\Big[\big[p,x\big]\,\hat{\mathcal{V}}_{0}(t_{1}-t_{2})\,\hat{x}_{\gamma_{2}}\,\hat{\mathcal{V}}(t_{2}-t')\big[x,\rho_{S}(t')\big]\Big]\nonumber\\& +(-i)^{2}\theta(t-t')\int_{t'}^{t} dt_{1} \int_{t'}^{t_{1}} dt_{2}\,\frac{\partial G^{r}_0(t-t_1)}{\partial t}\,\sum_{\gamma_{2}}\gamma_{2}\Sigma^{+\gamma_{2}}(t_{1}-t_{2})\mathrm{Tr}_{S}\Big[\big[x,x\big] \hat{\mathcal{V}}_{0}(t_{1}-t_{2})\,\hat{x}_{\gamma_{2}}\,\hat{\mathcal{V}}(t_{2}-t')\big[x,\rho_{S}(t')\big]\Big]. 
\end{align}
Here we have used the relation $\Sigma^{+\gamma_2}(t_1-t_2)=\Sigma^{-\gamma_2}(t_1-t_2)$ when $t_1\ge t_2$. We see that the last term is zero as $\big[x,x\big]=0$ and the second term can be simplified by putting $\big[p,x\big]=-i$. As the commutator of $p$ and $x$ is a $c$-number, it can be taken out of the trace and as a result, the further calculations become much simpler. Finally, the expression reduces to
\begin{align}
    &(-i)^{3}\theta(t-t')\int_{t'}^{t} dt_{1} \int_{t'}^{t_{1}} dt_{2}\,G_{0}^{r}(t-t_1)\sum_{\gamma_{2}}\gamma_{2}\Sigma^{+\gamma_{2}}(t_{1}-t_{2})\mathrm{Tr}_{S}\Big[\hat{\mathcal{V}}_{0}(t_{1}-t_{2})\,\hat{x}_{\gamma_{2}}\,\hat{\mathcal{V}}(t_{2}-t')\big[x,\rho_{S}(t')\big]\Big]\\
    &=-i\theta(t-t')\int_{t'}^{t} dt_{1} \int_{t'}^{t_{1}} dt_{2}\,G_{0}^{r}(t-t_1)\sum_{\gamma_{2}}\gamma_{2}\Sigma^{+\gamma_{2}}(t_{1}-t_{2})\mathrm{Tr}_{S}\Big[\hat{\mathcal{V}}_{0}(t_{1}-t_{2})\,\hat{x}_{\gamma_{2}}\,\hat{\mathcal{V}}(t_{2}-t')\big[x,\rho_{S}(t')\big]\Big].
\end{align}
Again substituting the form of bare propagator $\hat{\mathcal{V}}_0(t_1-t_2)$ and performing the summation over $\gamma_2$, we get,
\begin{align}
    &=-i\int_{t'}^{t} dt_{1} \int_{t'}^{t_{1}} dt_{2}\,G_{0}^{r}(t-t_1)\sum_{\gamma_{2}}\gamma_{2}\Sigma^{+\gamma_{2}}(t_{1}-t_{2})\mathrm{Tr}_{S}\Big[e^{-iH(t_1-t_2)}\,\hat{x}_{\gamma_{2}}\,\hat{\mathcal{V}}(t_{2}-t')\big[x,\rho_{S}(t')\big]e^{+iH(t_1-t_2)}\Big]\\
    &=-i\int_{t'}^{t} dt_{1} \int_{t'}^{t_{1}} dt_{2}\,G_{0}^{r}(t-t_1)\sum_{\gamma_{2}}\gamma_{2}\Sigma^{+\gamma_{2}}(t_{1}-t_{2})\mathrm{Tr}_{S}\Big[\hat{x}_{\gamma_{2}}\,\hat{\mathcal{V}}(t_{2}-t')\big[x,\rho_{S}(t')\big]\Big]\\
    &=-i\int_{t'}^{t} \!\!\!\!dt_{1}\!\! \int_{t'}^{t_{1}}\!\!\!\! dt_{2}\Big\{G_{0}^{r}(t\!-\!t_1)\Sigma^{<}(t_{1}\!-\!t_{2})\mathrm{Tr}_{S}\Big[x\,\hat{\mathcal{V}}(t_{2}\!-\!t')\big[x,\rho_{S}(t')\big]\Big]\!-\!G_{0}^{r}(t\!-\!t_1)\Sigma^{t}(t_{1}\!-\!t_{2})\mathrm{Tr}_{S}\Big[x\hat{\mathcal{V}}(t_{2}\!-\!t')\big[x,\rho_{S}(t')\big]\Big]\Big\}.
\end{align}
Using the Green's function identity $\Sigma^{t}(t_1-t_2)-\Sigma^{<}(t_1-t_2)=\Sigma^{r}(t_1-t_2)$, we obtain,
\begin{align}
    &\int_{t'}^{t} dt_{1} \int_{t'}^{t_{1}} dt_{2}\, G^{r}_{0}(t-t_1)\Sigma^{r}(t_1-t_2) \mathrm{Tr}_{S}\Big[x\hat{\mathcal{V}}(t_{2}-t')\big[x,\rho_{S}(t')\big]\Big]\\
    &=\int_{t'}^{t} dt_{1} \int_{t_0}^{t_{1}} dt_{2}\, G^{r}_{0}(t-t_1)\Sigma^{r}(t_1-t_2) G^{r}(t_2,t'). \label{second_term_final_step}
\end{align}
Time ordering $t>t_1>t_2>t'$ in Eq. ~\eqref{second_term_final_step} ensures that the lower limit of the $t_1$ integration can be extended to $t_0$ as for $t_1<t'$, the above expression will be zero because of causality. In a similar manner, by substituting the form of $\hat{\mathcal{V}}_0$ and performing the summations in Eq.~\eqref{Gr_dyson}, one can easily show that it is zero. Hence the Dyson equation of $G^{r}(t,t')$ becomes
\begin{align}
    G^{r}(t,t')=G_{0}^{r}(t,t')+\int_{t_0}^{t} dt_{1} \int_{t_0}^{t_{1}} dt_{2} G^{r}_{0}(t-t_1)\Sigma^{r}(t_1-t_2) G^{r}(t_2,t'). 
\end{align}
Remarkably, this is the exact Dyson equation of the retarded correlator $G^{r}(t,t')$. In summary, for the CL model, the Born approximated standard definition of the retarded correlator produces the exact Dyson equation for arbitrary time instants $t$ and $t'$. The reason behind this is that the commutator of $x_{I}$ in Eq.~\eqref{XI_CL} for two different times is a $c$-number which is purely because of the bilinear nature of the system Hamiltonian.
%
%
%
%
%
%
%
%
\section{General condition for getting no correction to the Born approximated retarded Green's function $G_{AB}^r(t,t')$ in $O(\lambda^{2})$}
\label{Gr_corr_zero}
In this appendix, we derive a general condition for getting no correction to the Born approximated retarded correlator. Hence, we start with obtaining the correction term for the lesser component of Green's function $G^{<}_{c,AB}(t,t')$ (for $t\ge t'$) by using the correction term of greater Green's function.
In Eq.~\eqref{general_cor} of the main text we have provided a correction term to the Born approximated definition for the greater component of the Green's function for $t\ge t'$. For $t\le t'$, a similar correction term can also be obtained by expanding the Born approximated greater component for $t\le t'$ in a similar manner like Eq.~\eqref{diagram_expansion} and following the same Feynman rules discussed earlier. Then one can easily identify the missing diagram corresponding to the finite system-bath correlation at time $t$. The mathematical expression of the correction term of greater Green's function $G^{>}_{AB}(t,t')$ for $t\le t'$ takes the form,
\begin{align}
    G^{>}_{c,AB}(t,t')= (-i)^2\xi\int_{t}^{t'}dt_{1}\int_{t_{0}}^{t} dt_{2}\sum_{\gamma_{1},\gamma_{2};i,j}\gamma_{1}\gamma_{2}\Sigma^{\gamma_{1}\gamma_{2}}_{ij}(t_{1}-t_{2})\mathrm{Tr}_{S}\Big[B\hat{\mathcal{V}}_{0}(t'-t_{1})\hat{S}^{i}_{\gamma_{1}}\hat{\mathcal{V}}_{0}(t_{1}-t)\Big(\hat{\mathcal{V}}(t-t_{2})\hat{S}^{j}_{\gamma_{2}}\rho_{S}(t_{2})\Big)A\Big].
\end{align}
Recall that $\xi=\pm 1$ based on our choice. Now using the following relation i.e., $G^{>}_{BA}(t',t)=G^{<}_{AB}(t,t')$, we write the similar correction term for $G^{<}_{AB}(t,t')$ for $t\ge t'$ as,
\begin{align}
    \theta(t-t')G^{<}_{c,AB}(t,t')&=\xi\,\theta(t-t')G^{>}_{c,BA}(t',t)\\
    &=(-i)^2\xi\!\int_{t'}^{t}\!\!\!dt_{1}\int_{t_{0}}^{t'} \!\!\!\!dt_{2}\!\!\!\sum_{\gamma_{1},\gamma_{2};i,j}\!\!\!\gamma_{1}\gamma_{2}\Sigma^{\gamma_{1}\gamma_{2}}_{ij}(t_{1}\!-\!t_{2})\mathrm{Tr}_{S}\Big[A\hat{\mathcal{V}}_{0}(t\!-\!t_{1})\hat{S}^{i}_{\gamma_{1}}\hat{\mathcal{V}}_{0}(t_{1}\!-\!t')\Big(\hat{\mathcal{V}}(t'\!-\!t_{2})\hat{S}^{j}_{\gamma_{2}}\rho_{S}(t_{2})\Big)B\Big].
\end{align}
As a result, the correction term for general retarded correlator $G^{r}_{c,AB}(t,t')$ (recall that the subscript $c$ here refers to the correction term) can be received from the following relation as,
\begin{align}
    G^{r}_{c,AB}(t,t')
    &=\theta(t-t') \Big[G^{>}_{c,AB}(t,t')-G^{<}_{c,AB}(t,t')\Big].
    \label{relation-GF}
\end{align}
In summary, the correction term of $G^{>}_{c,AB}(t,t')$
and $G^{<}_{c,AB}(t,t')$ for $t\ge t'$ obtained in the main text is the following,
\begin{align}
    &G_{c,AB}^{>}(t,t')= (-i)^2\int_{t'}^{t}\!\!\!dt_{1}\!\!\int_{t_{0}}^{t'}\!\!\! dt_{2}\!\!\sum_{\gamma_{1},\gamma_{2};i,j}\!\!\!\gamma_{1}\gamma_{2}\,\Sigma^{\gamma_{1}\gamma_{2}}_{ij}(t_{1}\!-\!t_{2})\mathrm{Tr}_{S}\Big[A\hat{\mathcal{V}}_{0}(t\!-\!t_{1})\hat{S}^{i}_{\gamma_{1}}\hat{\mathcal{V}}_{0}(t_{1}\!-\!t')B\hat{\mathcal{V}}(t'\!-\!t_{2})\hat{S}^{j}_{\gamma_{2}}\rho_{S}(t_{2})\Big],\\
    &G^{<}_{c,AB}(t,t')= (-i)^2\xi\int_{t'}^{t}\!\!\!dt_{1}\!\!\int_{t_{0}}^{t'}\!\!\!
    \!dt_{2}\!\!\!\sum_{\gamma_{1},\gamma_{2};i,j}\!\!\!\gamma_{1}\gamma_{2}\,\Sigma^{\gamma_{1}\gamma_{2}}_{ij}(t_{1}\!-\!t_{2})\mathrm{Tr}_{S}\Big[A\hat{\mathcal{V}}_{0}(t\!-\!t_{1})\hat{S}^{i}_{\gamma_{1}}\hat{\mathcal{V}}_{0}(t_{1}\!-\!t')\Big(\hat{\mathcal{V}}(t'\!-\!t_{2})\hat{S}^{j}_{\gamma_{2}}\rho_{S}(t_{2})\Big)B\Big].
\end{align}
Now let us first simplify the expression of $G^{>}_{c,AB}(t,t')$ by substituting the form of $\hat{\mathcal{V}}_0$ and performing the summations. 
\begin{align}
    G_{c,AB}^{>}(t,t')&= (-i)^2\int_{t'}^{t}\!\!\!dt_{1}\!\int_{t_{0}}^{t'}\!\!\! dt_{2}\!\!\sum_{\gamma_{1},\gamma_{2};i,j}\!\!\!\gamma_{1}\gamma_{2}\,\Sigma^{\gamma_{1}\gamma_{2}}_{ij}(t_{1}\!-\!t_{2})\mathrm{Tr}_{S}\Big[A\hat{\mathcal{V}}_{0}(t\!-\!t_{1})\hat{S}^{i}_{\gamma_{1}}\hat{\mathcal{V}}_{0}(t_{1}\!-\!t')B\hat{\mathcal{V}}(t'\!-\!t_{2})\hat{S}^{j}_{\gamma_{2}}\rho_{S}(t_{2})\Big]\\
    &=(-i)^2\int_{t'}^{t}\!\!\!dt_{1}\!\!\int_{t_{0}}^{t'}\!\!\! dt_{2}\!\!\sum_{\gamma_{1},\gamma_{2};i,j}\!\!\!\gamma_{1}\gamma_{2}\,\Sigma^{\gamma_{1}\gamma_{2}}_{ij}(t_{1}\!-\!t_{2})\mathrm{Tr}_{S}\Big[A_{I}(t\!-\!t_1)\hat{S}^{i}_{\gamma_{1}}\hat{\mathcal{V}}_{0}(t_{1}\!-\!t')B\hat{\mathcal{V}}(t'\!-\!t_{2})\hat{S}^{j}_{\gamma_{2}}\rho_{S}(t_{2})\Big]\\
    &=(-i)^2\int_{t'}^{t}\!\!\!dt_{1}\!\!\int_{t_{0}}^{t'}\!\!\! dt_{2}\!\!\sum_{\gamma_{2};i,j}\!\!\!\gamma_{2}\,\Sigma^{+\gamma_{2}}_{ij}(t_{1}\!-\!t_{2})\mathrm{Tr}_{S}\Big[\big[A_{I}(t\!-\!t'),S^{i}_{I}(t_1\!-\!t')\big]B\hat{\mathcal{V}}(t'\!-\!t_{2})\hat{S}^{j}_{\gamma_{2}}\rho_{S}(t_{2})\Big].
\end{align}
In a similar way, we will simplify the correction term of the lesser component $G^{<}_{c,AB}(t,t')$ as the following,
\begin{align}
    G^{<}_{c,AB}(t,t')&= (-i)^2\xi\int_{t'}^{t}\!\!\!dt_{1}\!\!\int_{t_{0}}^{t'} \!\!\!\!dt_{2}\!\!\!\sum_{\gamma_{1},\gamma_{2};i,j}\!\!\!\gamma_{1}\gamma_{2}\,\Sigma^{\gamma_{1}\gamma_{2}}_{ij}(t_{1}\!-\!t_{2})\mathrm{Tr}_{S}\Big[A\hat{\mathcal{V}}_{0}(t\!-\!t_{1})\hat{S}^{i}_{\gamma_{1}}\hat{\mathcal{V}}_{0}(t_{1}\!-\!t')\Big(\hat{\mathcal{V}}(t'\!-\!t_{2})\hat{S}^{j}_{\gamma_{2}}\rho_{S}(t_{2})\Big)B\Big]\\
    &=(-i)^2\xi\int_{t'}^{t}\!\!\!dt_{1}\!\!\int_{t_{0}}^{t'}\!\!\!\! dt_{2}\!\!\sum_{\gamma_{1},\gamma_{2};i,j}\!\!\!\gamma_{1}\gamma_{2}\!\!\!\Sigma^{\gamma_{1}\gamma_{2}}_{ij}(t_{1}\!-\!t_{2})\mathrm{Tr}_{S}\Big[A_{I}(t\!-\!t_{1})\hat{S}^{i}_{\gamma_{1}}\hat{\mathcal{V}}_{0}(t_{1}\!-\!t')\Big(\hat{\mathcal{V}}(t'\!-\!t_{2})\hat{S}^{j}_{\gamma_{2}}\rho_{S}(t_{2})\Big)B\Big]\\
    &=(-i)^2\xi\int_{t'}^{t}\!\!\!dt_{1}\!\!\int_{t_{0}}^{t'} \!\!\!\!dt_{2}\!\!\!\sum_{\gamma_{2};i,j}\!\!\!\gamma_{2}\,\Sigma^{+\gamma_{2}}_{ij}(t_{1}\!-\!t_{2})\mathrm{Tr}_{S}\Big[\big[A_{I}(t\!-\!t'),S^{i}_{I}(t_1\!-\!t')\big]\Big(\hat{\mathcal{V}}(t'\!-\!t_{2})\hat{S}^{j}_{\gamma_{2}}\rho_{S}(t_{2})\Big)B\Big]\\
    &=(-i)^2\xi\int_{t'}^{t}\!\!\!dt_{1}\!\!\int_{t_{0}}^{t'} \!\!\!\!dt_{2}\!\!\!\sum_{\gamma_{2};i,j}\!\!\!\gamma_{2}\Sigma^{+\gamma_{2}}_{ij}(t_{1}\!-\!t_{2})\mathrm{Tr}_{S}\Big[B\big[A_{I}(t\!-\!t'),S^{i}_{I}(t_1\!-\!t')\big]\hat{\mathcal{V}}(t'\!-\!t_{2})\hat{S}^{j}_{\gamma_{2}}\rho_{S}(t_{2})\Big].
\end{align}
Following the relation in Eq.~\eqref{relation-GF}, finally the correction to the retarded correlator $G^{r}_{c,AB}(t,t')$ becomes,
\begin{align}
    G^{r}_{c,AB}(t,t')=(-i)^2\xi\int_{t'}^{t}\!\!\!dt_{1}\!\!\int_{t_{0}}^{t'} \!\!\!\!dt_{2}\!\!\!\sum_{\gamma_{2};i,j}\!\!\!\gamma_{2}\,\Sigma^{+\gamma_{2}}_{ij}(t_{1}\!-\!t_{2})\mathrm{Tr}_{S}\Big\{\Big[\big[A_{I}(t\!-\!t'),S^{i}_{I}(t_1\!-\!t')\big],B\Big]_{\xi}\hat{\mathcal{V}}(t'\!-\!t_{2})\hat{S}^{j}_{\gamma_{2}}\rho_{S}(t_{2})\Big\}.
\end{align}
As discussed in the main text, for $\xi=+1$, the expression $\Big[\big[A_{I}(t-t'),S^{i}_{I}(t_1-t')\big],B\Big]_{\xi}$ becomes a commutator $\Big[\big[A_{I}(t-t'),S^{i}_{I}(t_1-t')\big],B\Big]$ and for $\xi=-1$, it becomes an anti-commutator $\Big\{\big[A_{I}(t-t'),S^{i}_{I}(t_1-t')\big],B\Big\}$.
We finally obtain the condition for the correction to the Born approximated retarded correlator to be zero  and it is given by,
\begin{equation}
    \Big[\big[A_{I}(t-t'),S^{i}_{I}(t_1-t')\big],B\Big]_{\xi}=0. \label{retarded_corr_zero_cond}
\end{equation}
A similar condition can be arrived at for the advanced component and is given as,
\begin{equation}
    \Big[\big[B_{I}(t'-t),S^{i}_{I}(t_1-t)\big],A\Big]_{\xi}=0. \label{retarded_corr_zero_cond}
\end{equation}
%
%
%
\subsection{Absence of correction to the Born approximated retarded correlator for the Caldeira-Leggett model to $O(\lambda^2)$}
For Caldeira-Leggett (CL) model, the operators $A$, and $B$ are both the position operator $x$ and there is only one system operator coupled to the bath which is also $x$, hence we will write $S^{j}=x$. Now, we will check whether the condition for the correction to the retarded Green's function to be zero is satisfied. Here our choice of $\xi$ is $+1$. We obtain,
\begin{align}
    &\Big[\big[x_{I}(t-t'),x_{I}(t_1-t')\big],x\Big]\\
    &=\Big[\Big(G^{r}_{0}(t_1-t)F^{r}_{0}(t-t')\big[x,p\big]+G^{r}_{0}(t-t')F^{r}_{0}(t_1-t')\big[p,x\big]\Big),x\Big]\\
    &=-i\Big[\Big(G^{r}_{0}(t_1-t)F^{r}_{0}(t-t')-G^{r}_{0}(t-t')F^{r}_{0}(t_1-t')\Big),x\Big]\\
    &=0.
\end{align}
Note that the absence of correction to $O(\lambda^2)$ for the CL model is because of the bilinear structure in the Hamiltonian and hence the commutator $\big[x_{I}(t-t'),x_{I}(t_1-t')\big]$ is a $c$-number. In fact, in Appendix \ref{exact_dyson_Gr} we show that, because of this non-interacting nature of CL model, the Born approximated retarded Green's function predicts the exact Dyson equation.
\subsection{Absence of correction to the retarded correlator for the spin-boson model to $O(\lambda^2)$}
For the spin-boson (SB) model, we analyze two different components for the retarded correlator, i.e., $G^{r}_{xx}(t,t')$ and $G^{r}_{yx}(t,t')$. As here our choice of $\xi$ is $-1$, hence the $[\,\,,\,\,]_\xi$ is an anti-commutator. For the SB model, $S$ operator is $\sigma_x$. We therefore need the expressions of $\sigma_{x,I}(t)$ and $\sigma_{y,I}(t)$ which is the following,
\begin{align}
    &\sigma_{x,I}(t)=\sigma_x \mathrm{cos}\,\omega_0 t -\sigma_y \mathrm{sin}\,\omega_0 t, \\
    &\sigma_{y,I}(t)=\sigma_x \mathrm{sin}\,\omega_0 t + \sigma_y \mathrm{cos}\,\omega_0 t.
\end{align}
Now let us check the condition defined in Eq~\eqref{retarded_corr_zero_cond} for the $G^{r}_{c,xx}(t,t')$ correlator.
\begin{align}
    &\Big\{\big[\sigma_{x,I}(t-t'),\sigma_{x,I}(t_1-t')\big],\sigma_x\Big\}\\
    &=\Big\{\Big(\mathrm{cos}\,\omega_0(t_1-t')\mathrm{sin}\,\omega_0(t-t')[\sigma_x,\sigma_y]-\mathrm{cos}\,\omega_0(t-t')\mathrm{sin}\,\omega_0(t_1-t')[\sigma_x,\sigma_y]\Big),\sigma_x\Big\}\\
    &=2i\,\mathrm{sin}\,\omega_0(t-t_1)\Big\{\sigma_z,\sigma_x\Big\}\\
    &=0.
\end{align}
Similarly we can check the condition for the $G^{r}_{c,yx}(t,t')$ which also appears to be  zero,
\begin{align}
    &\Big\{\big[\sigma_{y,I}(t-t'),\sigma_{x,I}(t_1-t')\big],\sigma_x\Big\}\\
    &=-\Big\{\Big(\mathrm{sin}\,\omega_0(t_1-t')\mathrm{sin}\,\omega_0(t-t')[\sigma_x,\sigma_y]-\mathrm{cos}\,\omega_0(t-t')\mathrm{cos}\,\omega_0(t_1-t')[\sigma_x,\sigma_y]\Big),\sigma_x\Big\}\\
    &=2i\,\mathrm{cos}\,\omega_0(t-t_1)\Big\{\sigma_z,\sigma_x\Big\}\\
    &=0.
\end{align}
Hence the correction of $O(\lambda^{2})$ is zero for both the two retarded Green's function in spin-boson model. In a similar way, one can show that all the other components of the retarded Green's function will also satisfy the condition in Eq.~\eqref{retarded_corr_zero_cond} and hence does not need any correction in $O(\lambda^2)$.
\end{widetext}
 \bibliographystyle{apsrev4-1}
 \bibliography{references.bib}
\end{document}